\documentclass[twocolumn,tighten]{aastex63}
\usepackage{amsmath}
\usepackage{threeparttable}
\usepackage{float}
\usepackage{graphicx}
\usepackage{CJK}
\usepackage{xspace}

\defcitealias{SurveyPaper}{Paper I}

\submitjournal{ApJ}

\shorttitle{EIGER II: Characterisation of [OIII] emitters at $z\sim6$}
\shortauthors{Matthee et al.}

\begin{document}
\begin{CJK*}{UTF8}{gbsn}

\title{EIGER II. first spectroscopic characterisation of the young stars and ionised gas associated with strong H$\beta$ and [OIII] line-emission in galaxies at $z=5-7$ with {\it JWST} }

\correspondingauthor{Jorryt Matthee}
\email{mattheej@phys.ethz.ch}
\author[0000-0003-2871-127X]{Jorryt Matthee}
\affiliation{Department of Physics, ETH Z{\"u}rich, Wolfgang-Pauli-Strasse 27, Z{\"u}rich, 8093, Switzerland}

\author[0000-0003-0417-385X]{Ruari Mackenzie}
\affiliation{Department of Physics, ETH Z{\"u}rich, Wolfgang-Pauli-Strasse 27, Z{\"u}rich, 8093, Switzerland}

\author[0000-0003-3769-9559]{Robert A.~Simcoe}
\affiliation{MIT Kavli Institute for Astrophysics and Space Research, 77 Massachusetts Avenue, Cambridge, 02139, Massachusetts, USA}

\author[0000-0001-9044-1747]{Daichi Kashino}
\affiliation{Institute for Advanced Research, Nagoya University, Nagoya 464-8601, Japan}
\affiliation{Division of Particle and Astrophysical Science, Graduate School of Science, Nagoya University, Nagoya 464-8602, Japan}

\author[0000-0002-6423-3597]{Simon J.~Lilly}
\affiliation{Department of Physics, ETH Z{\"u}rich, Wolfgang-Pauli-Strasse 27, Z{\"u}rich, 8093, Switzerland}

\author[0000-0002-3120-7173]{Rongmon Bordoloi}
\affiliation{Department of Physics, North Carolina State University, Raleigh, 27695, North Carolina, USA}

\author[0000-0003-2895-6218]{Anna-Christina Eilers}
\affiliation{MIT Kavli Institute for Astrophysics and Space Research, 77 Massachusetts Avenue, Cambridge, 02139, Massachusetts, USA}

\begin{abstract}
We present emission-line measurements and physical interpretations for a sample of 117 [OIII] emitting galaxies at $z=5.33-6.93$, using the first deep {\it JWST}/NIRCam wide field slitless spectroscopic observations. Our 9.7-hour integration is centered upon the $z=6.3$ quasar J0100+2802 -- the first of six fields targeted by the EIGER survey -- and covers $\lambda=3-4$ microns. We detect 133 [OIII] doublets, but close pairs motivated by their small scale clustering excess. The galaxies are characterised by a UV luminosity M$_{\rm UV}\sim-19.6$ ($-17.7$ to $-22.3$), stellar mass $\sim10^8$ $(10^{6.8-10.1})$ M$_{\odot}$, H$\beta$ and [OIII]$_{4960+5008}$ EWs $\approx$ 850 {\AA} (up to 3000 {\AA}), young ages, a highly excited interstellar medium and low dust attenuations. These high EWs are very rare in the local Universe, but we show they are ubiquitous at $z\sim6$ based on the measured number densities. The stacked spectrum reveals H$\gamma$ and [OIII]$_{4364}$ which shows that the galaxies are typically dust and metal poor (E$(B-V)=0.1$, $12+\log(\mathrm{O/H})=7.4$) with a high electron temperature ($2\times10^4$ K) and a production efficiency of ionising photons ($\xi_{\rm ion}=10^{25.3}$ Hz erg$^{-1}$). We further show the existence of a strong mass-metallicity relation. The properties of the stars and gas in $z\sim6$ galaxies conspire to maximise the [OIII] output from galaxies, yielding an [OIII] luminosity density at $z\approx6$ that is significantly higher than at $z\approx2$. Thus, [OIII] emission-line surveys with {\it JWST} prove a highly efficient method to trace the galaxy density in the Epoch of Reionization.
\end{abstract}


\section{Introduction}
\label{sec:introduction}

In the last two decades, extensive observations with the {\it Hubble Space Telescope (HST)} and ground-based telescopes have revealed the first glimpse of galaxies in the early Universe at redshifts beyond the peak of cosmic star formation history ($z\gtrsim2$; \citealt{madau&dickinson}). Most emphasis has been on mapping out the evolution of the number density distribution of Lyman-Break galaxies \citep[LBGs; e.g.][]{Bunker10,Finkelstein16,Bouwens21}. These galaxies are selected through a sharp color-drop in photometric data that is particularly strong at high-redshift due to absorption from intervening neutral hydrogen \citep[e.g.][]{Madau95,Steidel96}. This is an effective technique that exploits the sensitive imaging capabilities on board the {\it HST}, but it is not very specific as samples contain interlopers and the estimated redshifts are not very accurate (see \citealt{Brinchmann17} for a detailed comparison between spectroscopic and photometric redshifts at $z\lesssim6$).

Observations suggest that the cosmic star formation rate density increased significantly by a factor $\sim10$ between $z\sim7$ and $z\sim2$. The average properties of LBGs evolve significantly over this time-interval, with galaxies at higher redshifts being bluer and smaller at fixed luminosity \citep[e.g.][]{Bouwens14,Shibuya15}. A key feature is that galaxies in the early Universe appear characterised by extremely strong nebular emission-lines, in part due to their increasing specific star formation rates (sSFR; e.g. \citealt{Stark13,MarmolQueralto16}). The evidence for strong nebular H$\beta$+[OIII] emission is the large color excesses in {\it Spitzer}/IRAC data covering the rest-frame optical \citep[e.g.][]{Zackrisson08,Schaerer09,Raiter10a,Labbe13,Smit14,Roberts-Borsani16a} and circumstantially the increasing fraction of strong Lyman-$\alpha$ emitters among LBGs with increasing redshift \citep[e.g.][]{Stark11,Sobral18,Kusakabe20}. However, the presence of these lines in large statistical samples of $z\sim6$ galaxies has not yet been confirmed spectroscopically.

Galaxies with extreme emission lines (EELGs; e.g. \citealt{vanderWel11}) are present throughout the history of the Universe \citep[e.g.][]{Izotov18b,Izotov21}. For example, Mrk 71 at a redshift of 80 km s$^{-1}$ ($z=0.0003$) has a rest-frame [OIII]$_{4960,5008}$ equivalent width (EW$_0$) of 864 {\AA} \citep{Moustakas2006,Micheva17}. However, galaxies with such extreme EWs $\approx1000$ {\AA} are very rare in the low redshift Universe. At stellar masses $\sim10^8$ M$_{\odot}$, the typical [OIII] EW is 20 {\AA} in the Sloan Digital Sky Survey (SDSS, e.g. \citealt{Alam15}), and only $<1$ \% of SDSS galaxies are an EELG. The typical EWs increase with redshift and are 100-200 {\AA} at $z\sim2$ \citep[e.g.][]{Khostovan2016,Malkan2017,Boyett21}, with a strong increase towards lower mass \citep{Reddy18b}. Photometric inferences beyond $z>3$ suggest typical EWs of $\approx700$ {\AA} at $z\sim6-8$ with similar mass dependencies \citep[e.g.][]{Labbe13,debarros19,Endsley21,Endsley22b}.

The physical conditions in the interstellar medium (ISM) that is associated to these young star-bursting systems is relatively unexplored at $z>3$, in particular for low mass galaxies. ALMA spectroscopy has revealed detections of strong 88 $\mu$m [OIII] line emission in a handful of galaxies at $z\sim7$ \citep[e.g.][]{Inoue16,Tamura19,Hashimoto21,Witstok22} showing high [OIII]/[CII] line-ratios suggestive of a high ionisation parameter \citep[$\approx5$;
][]
{Harikane20}. The {\it JWST} is now set to transform this field by enabling sensitive rest-frame optical spectroscopy at $z\approx3-9$. The first commissioning data have already revealed strong [OIII] line emission at $z=4-9$ in a handful of objects \citep{Rigby22,Sun22,Sun22b} and various other faint lines suggestive of very young ages, low metallicities and a highly ionised ISM \citep[e.g.][]{Brinchmann22,Carnall22,Curti22,Katz22,Rhoads22,Schaerer22,Tacchella22b,Taylor22,Trump22}. Large surveys of statistical samples are soon anticipated \citep[e.g.][]{Bunker20}.

In this paper we present the characterisation of a sample of 117 spectroscopically confirmed [OIII] doublets at $z=5.33-6.93$ from the first observations of our large {\it JWST} EIGER program (Emission-line galaxies and the Intergalactic Gas in the Epoch of Reionization; program ID 1243, PI Lilly). As explained and motivated in detail in our survey paper \citep[][hereafter Paper I]{SurveyPaper}, EIGER uses wide-field slitless spectroscopy (WFSS) with NIRCam to obtain complete samples of H$\alpha$ and [OIII] emission line galaxies at $z=3-7$ in the fields of six bright quasars at $z=6-7$. The main goal of EIGER is to study the end stages of cosmic reionization and the metal enriched envelopes of early galaxies through cross-correlations between galaxies and hydrogen and metal absorption lines.

The survey design, observing strategy and the first cross-correlation between galaxies and the Ly$\alpha$ forest at $z\sim6$ in the field of the ultra-luminous quasar J0100+2802 at $z=6.33$ \citep{Wu15,Wang16} are presented in \citetalias{SurveyPaper}. Here we will briefly present the observations and data reduction in \S $\ref{sec:data}$. In \S $\ref{sec:methods}$, we present the techniques to identify emission-line galaxies in the grism data, we motivate a method to merge closely separated clumps into systems and present the method to measure line-fluxes and model the spectral energy distribution (SEDs) of our data. In \S $\ref{sec:stronglines}$ we show the first spectroscopic evidence for strong rest-frame optical line emission in a large sample of galaxies at $z\sim6$ and measure the [OIII] luminosity function. Then we present the physical conditions (ionising photon production efficiency, dust attenuation, gas-phase metallicity) in our sample of [OIII] emitters based on spectroscopic measurements in \S $\ref{sec:physcond}$. Our results are discussed in \S $\ref{sec:discuss}$ and we summarise our results and their interpretation in \S $\ref{sec:summary}$. 

Throughout this work we assume a flat $\Lambda$CDM cosmology with $H_0=67.4$ km s$^{-1}$ Mpc$^{-1}$ and $\Omega_M=0.315$ \citep{Planck18}. Magnitudes are listed in the AB system.

\section{Data}
\label{sec:data}
The details of the survey design and data reduction are presented in \citetalias{SurveyPaper}. Here we briefly summarise these and highlight the aspects that are most relevant for this paper.

\begin{figure*}
    \centering
    \includegraphics[width=18cm]{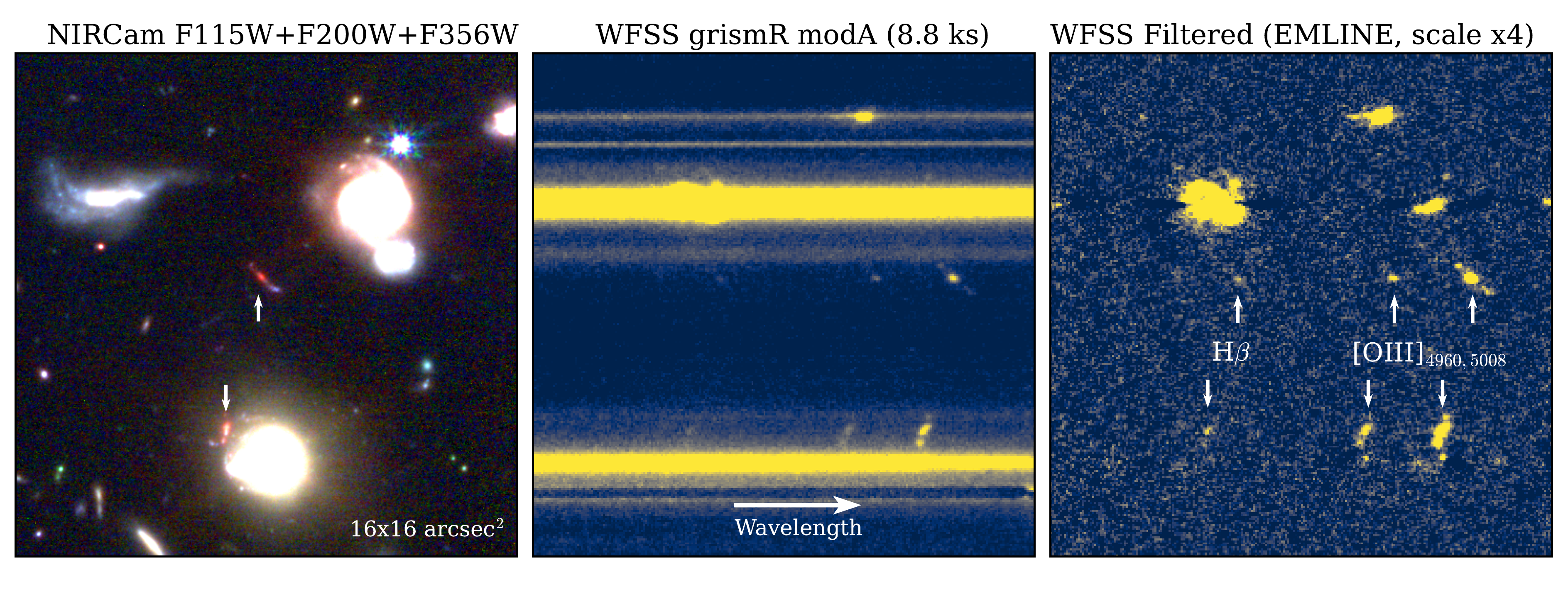}
    \caption{Demonstration of the {\it JWST}/NIRCam imaging and grism data and the continuum-filtering efficiency in a small $16\times16$ arcsec$^2$ region that constitutes 0.3 \% of the data in the J0100+2802 field.  The left panel shows a false-color composite of the F115W/F200W/F356W imaging and highlights the locations of two [OIII] emitting systems identified in our data. These are particularly red due to the strong line-emission that falls in the F356W filter. The middle panel shows the dispersed grism image on the same sub-region, while the right panel shows the result of our continuum-filtering methodology which reveals various emission-lines detected in the data. 
    }
    \label{fig:data_illustration}
\end{figure*}

\subsection{Observations}
We use a combination of infrared imaging and wide-field slitless spectroscopy (WFSS) of the high-redshift quasar J0100+2802 taken with NIRCam \citep{Rieke22} on the {\it JWST} (program ID 1243, PI Lilly). The spectroscopic component consists of grism integrations in the F356W filter using GRISMR that disperses spectra in the horizontal direction. The imaging data consist of F115W and F200W (short-wave channel) observations taken contemporaneously with the spectroscopy, and direct and out-of-field imaging in the F356W filter that covers the spectroscopic field of view. As detailed in \citetalias{SurveyPaper}, we employ a four pointing mosaic centered on the quasar. A central region of about $40''\times40''$ is observed during all four visits, with several further regions that are covered by two or one visits. The total area with spectroscopic coverage is 25.9 arcmin$^2$, of which $\approx4.6$ arcmin$^2$ is covered by both NIRCam modules A and B (with reversed dispersion direction). The observations were undertaken on four visits on 22-24 August 2022 with a position angle of the pointing of 236 degrees. The total grism exposure times range from 8.8-35.0 ks, whereas the direct imaging time ranges from 1.6-6.3 ks and the imaging in the short wavelengths ranges from 4.4-23.8 ks, with the F200W imaging receiving about 35 \% more exposure time than the F115W imaging.
\
\subsection{Imaging data reduction and photometry} \label{sec:imaging}
As detailed in Paper I, the NIRCam imaging data are reduced based on a combination of the \texttt{jwst} pipeline v-1.8.2\footnote{https://github.com/spacetelescope/jwst}\footnote{We used the CRDS context \texttt{jwst$\_$0988.pmap}, released on 25 October 2022 on the PUB server. This uses zero-points based on in-flight data \citep{Boyer22}.} and additional post-processing procedures. We perform the standard steps from \texttt{Detector1} and \texttt{Image2} and aligned the images to a common astrometric reference system aligned to {\sc Gaia} \citep{Gaia18}. We perform additional subtraction of the sky level and stray-light features (`wisps') and masked large residual cosmic-ray features \citep[`snowballs' e.g.][]{Merlin22}. Using a deep source mask we filter the 1/f noise in our exposures, inspired by \cite{Schlawin20} we subtract the median sky in quarter rows, then columns and finally in the four amplifiers. \texttt{Image3} is used to combine the post-processed images onto a common grid with resolution 0.03$''$/pixel.

Aperture-matched photometry is performed using \texttt{SExtractor} \citep{Bertin96} with the F356W imaging data as detection image. The higher resolution F115W and F200W imaging data are convolved to match the point spread function (PSF) of the F356W imaging \citepalias[see][]{SurveyPaper}. Magnitudes are measured with Kron apertures and the errors estimated from the random blank sky variation for apertures of different sizes, scaled to the local variance propagated by the pipeline \citep[following ][]{Finkelstein22}. The typical 5$\sigma$ sensitivities are 27.8, 28.3, 28.1 in the F115W, F200W and F356W imaging data, respectively, reaching to a magnitude deeper in the best regions \citepalias[see][]{SurveyPaper}.

\subsection{WFSS data reduction} 
WFSS data are reduced with a combination of the \texttt{jwst} pipeline (version 1.7.0) and our \texttt{python} based processing steps as detailed in Paper I, which we summarise here. Each exposure is processed with \texttt{Detector1} step and assigned a WCS using \texttt{Spec2}. \texttt{Image2} is used for flat-fielding and we additionally remove $1/f$ noise and sky background variations by subtracting the median value in each column. The output after this step is named `SCI'. Our main development is the separation of the `SCI' image in two components, `EMLINE' and `CONT', which separate emission-lines from the continuum. The continuum filtering subtracts the running median in the dispersion direction with a filter with a flexible kernel that has a hole in the center not to over-subtract lines themselves. The process does not rely on the trace model or known positions of continuum sources, but does also not distinguish between continuum from sources themselves, or contamination. We illustrate the efficiency of the continuum filtering methodology in Fig. $\ref{fig:data_illustration}$. Despite that one of the (multiple component) [OIII] emitters is very close to a continuum bright galaxy, that spectral trace is not visible in the EMLINE image.

Spectral extraction for each object detected in the imaging (\S $\ref{sec:imaging}$) is performed based on \texttt{grismconf}\footnote{https://github.com/npirzkal/GRISMCONF} using the latest (V4) trace models from N. Pirzkal, F. Sun and E. Egami\footnote{https://github.com/npirzkal/GRISM$\_$NIRCAM}. We have verified the accuracy of the trace model for GRISMR in both modules in the F356W filter using extracted spectra of faint (F356W$\sim20$) stars in our data (see also \citealt{Sun22}) and apply pixel-level corrections when necessary. We extract 2D spectra from the SCI, EMLINE, CONT and ERR extensions in each of the (at max) 96 grism images. These are divided by the relevant sensitivity curve (hence correcting for different sensitivities in modules A and B), rectified for small curvature of the trace and scrunched to a common observed wavelength grid (from 3.0 to 4.0 $\mu$m in steps of 9.75 {\AA}). We then create stacked mean spectra which were 5$\sigma$-clipped in three iterations. In addition, we create separate stacks for subsets of the individual visits and modules. We find that the spectra of sources observed in multiple visits (i.e. on different parts of the detector and/or on different modules) align excellently and have consistent fluxes.

\begin{figure}
    \centering
    \includegraphics[width=8.1cm]{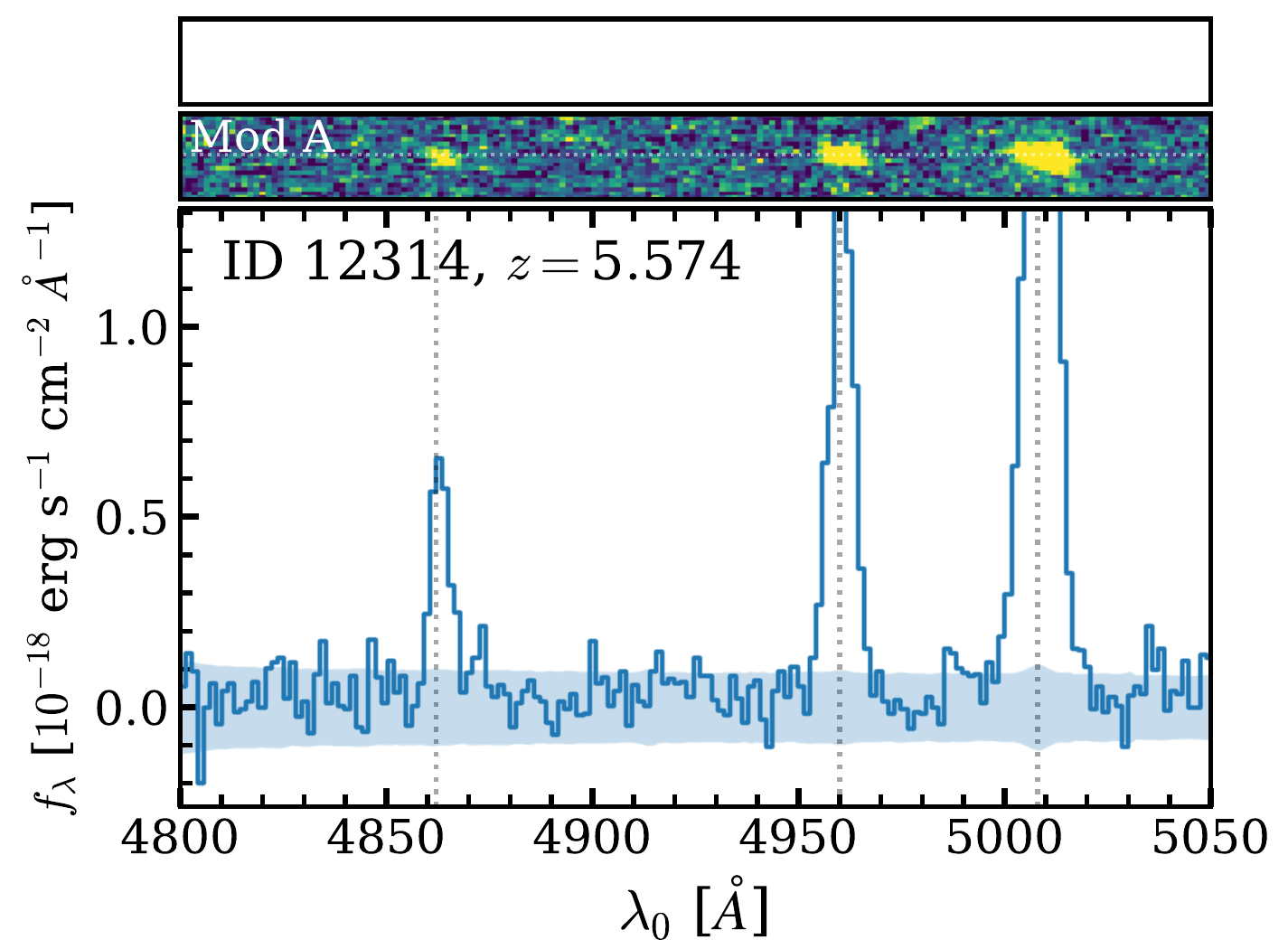} \\
        \includegraphics[width=8.1cm]{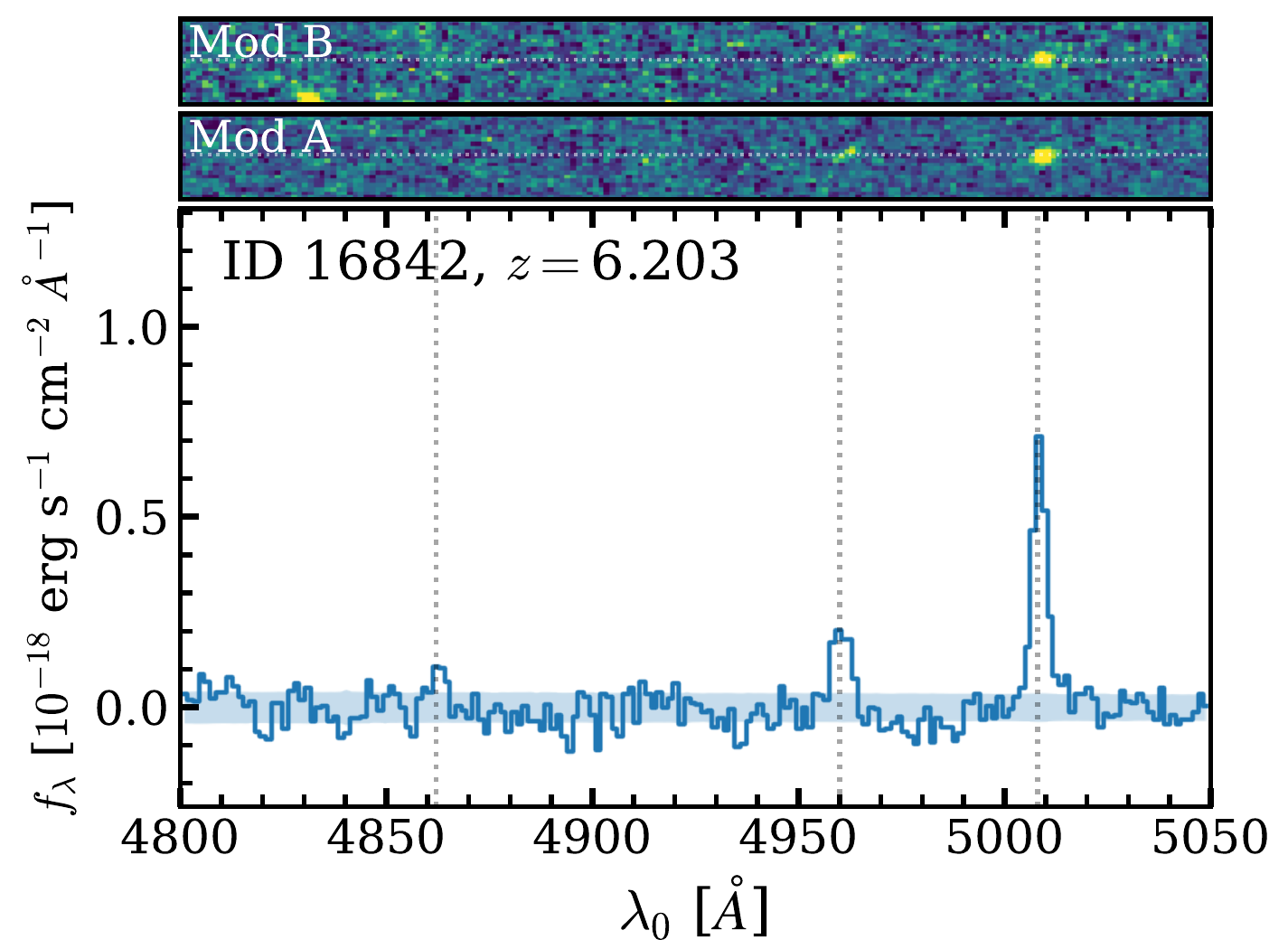} \\
    \includegraphics[width=8.1cm]{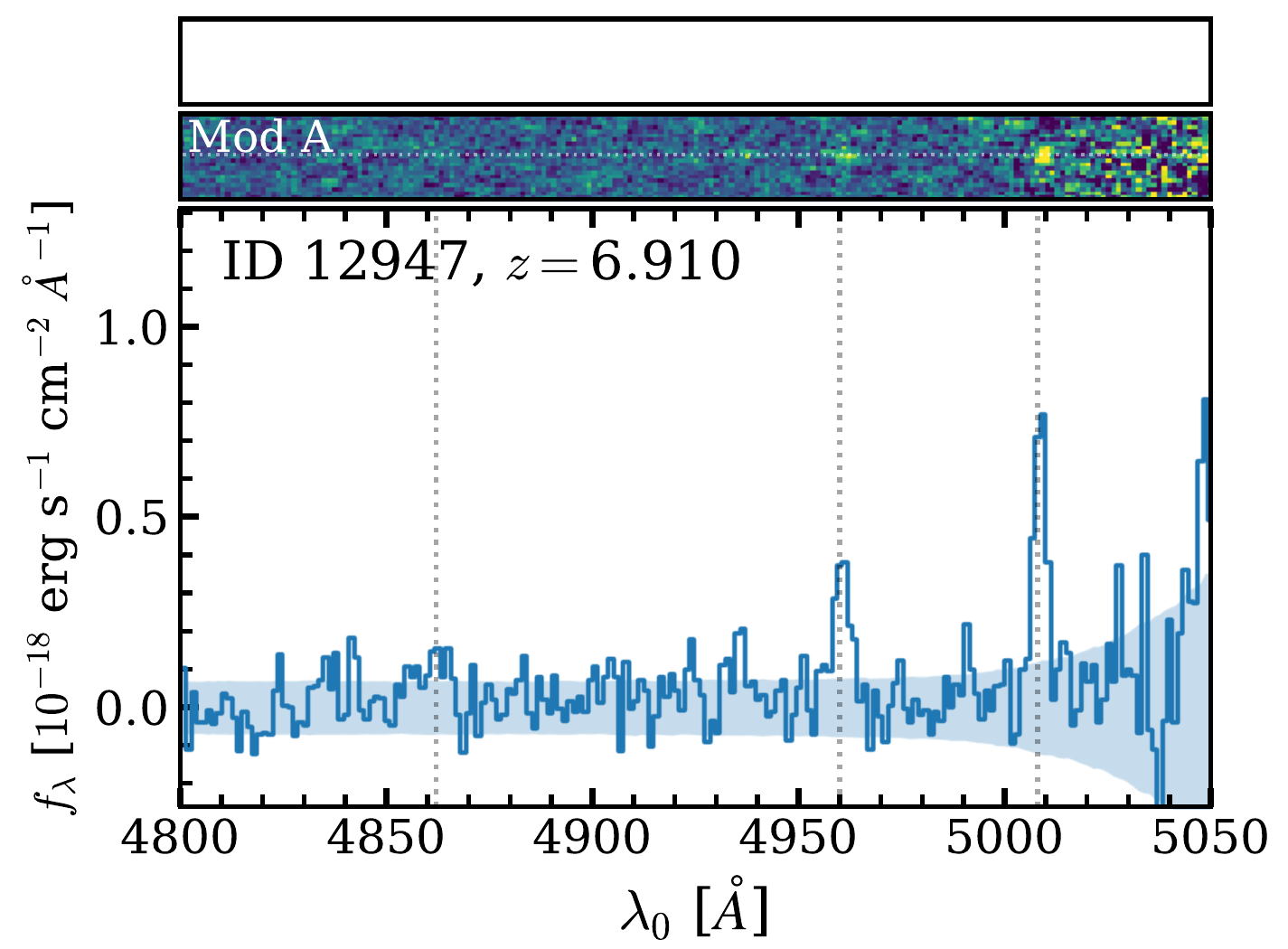} \\ 
    \caption{Example emission-line 1D spectra of three representative [OIII] emitters in our sample. Vertical dotted lines highlight the locations of H$\beta$ and [OIII]$_{4960,5008}$. Shaded regions show the noise level. The integrated S/N of H$\beta$ ([OIII]$_{4960}$) in each panel are  13.8 (38.3), 4.5 (10.8) and 3.3 (7.4), respectively. }
    \label{fig:example_spectra}
\end{figure}

\begin{figure*}
\begin{tabular}{ccc}
\hspace{-0.4cm}
 \includegraphics[width=5.3cm]{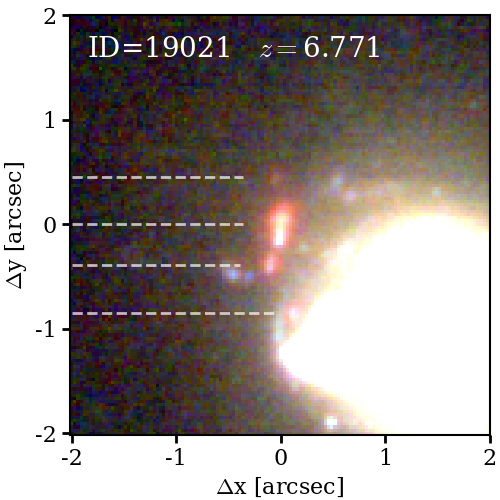}    & 
  \includegraphics[width=5.3cm]{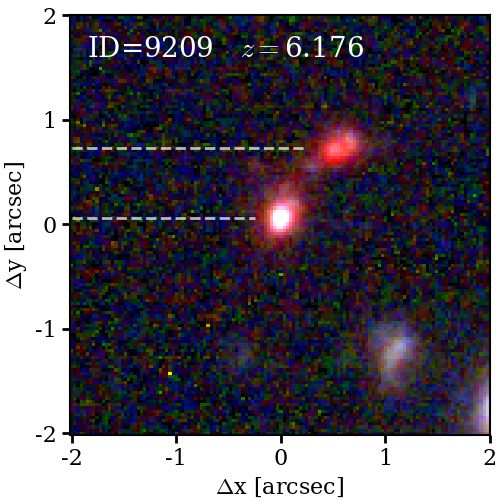}    & 
   \includegraphics[width=5.3cm]{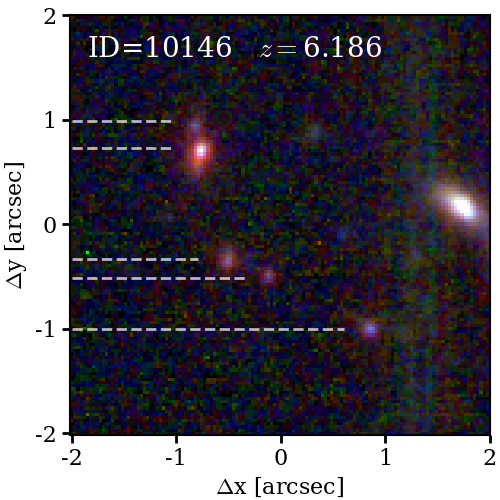}    \\ 
\end{tabular}
    \caption{Example false-color F115W/F200W/F356W stamps of regions where we detect multiple [OIII] emitting systems within 2$''$, highlighting the diversity of these groups. The images are oriented with the position angle 236 degrees. Each horizontal dashed line marks an [OIII] emitting component that is resolved in the grism data. }
    \label{fig:collage}
\end{figure*} 

\section{Identification of [OIII] systems} \label{sec:methods}
\subsection{Detection algoritms} 
As detailed in Paper I, we identify [OIII] emitters with two complementary approaches, named the ``backward" and the ``forward" approach. In the backward approach, we identify emission-line pairs (such as [OIII]$_{4960,5008}$) or triplets (e.g. with H$\beta$) by running \texttt{SExtractor} directly on the combined continuum-filtered grism images that are stacked per visit and module, and then identify the corresponding galaxy on the direct image based on the observed wavelength that is estimated from the observed doublet separation. In the forward approach, we run \texttt{SExtractor} on the direct F356W image, and then extract a spectrum in the EMLINE image for each galaxy at its expected position based upon the tracing model, identifying galaxies for which candidate pairs of lines are detected close to the expected trace center. Lines are detected with a minimum signal to noise ratio (S/N) of 3. The 3$\sigma$ limiting sensitivity of the spectroscopic data varies across the field and with wavelength (by a factor $\approx2$), with the best sensitivity of $0.6\times10^{-18}$ erg s$^{-1}$ cm$^{-2}$ at 3.8 $\mu$m. In order to facilitate the line - galaxy association and the redshift identification of each of the lines we limit ourselves to objects with at least two significant line detections.

In total, after careful visual inspections and reconciliation of the objects identified with the two methodologies and an iterative fine-tuning of the search parameters, we identify 133 resolved [OIII] emitting components over $z=5.33-6.93$ with at least two detected emission-lines with S/N$>3$. Fig. $\ref{fig:example_spectra}$ shows three example spectra that are representative for the full sample. All H$\beta$+[OIII] spectra are shown in Appendix $\ref{appendix:1D}$. The typical S/N for the bright [OIII]$_{5008}$ line is 14 (ranging from 6-70) and H$\beta$ is detected with S/N$>3 (5)$ in 68 (31) objects (detectable at $z\gtrsim5.5$ in our data). Only 3/133 objects were identified thanks to H$\beta$ (i.e. [OIII]$_{4960}$ S/N$<3$). We detect H$\gamma$ in two objects with S/N=3.8 and 7.6, respectively. The catalog of [OIII] emitters including their coordinates, confidence flags and redshifts will be released with our survey paper \citepalias{SurveyPaper}.

\subsection{Definition of a system}
While inspecting the [OIII] emitters, we noticed that a significant number appear in closely separated pairs or multiples (see also \citealt{Chen22} for similar results based on {\it JWST} imaging). Here, we argue for a simple definition of a `system' that can be easily mimicked in simulations.

Figure $\ref{fig:collage}$ shows false color images of example [OIII] emitters with multiple closely separated components. The object in the left panel is resolved in four components within 0.4 arcsec ($\sim2$ kpc) while being at a close distance to a foreground galaxy suggesting possible galaxy-galaxy lensing (see also Fig. $\ref{fig:data_illustration}$). Several such multiples exist in our sample, but such a close separation to the foreground galaxy is rare. The object in the middle panel shows two clumps with comparable luminosity and faint emission bridging these components. The right panel shows a group of five galaxies with a somewhat larger separation (maximally $2.5''$ or $\approx15$ kpc). These five galaxies are identified as four clumps in our source-catalog. 

This simple compilation of galaxies in our data suggests that we should adopt a physical criterion to define an individual system, instead of relying on the specific choice of deblending parameters used by \texttt{SExtractor}, the orientation of the multiple components, or the specific spatial resolution of the data. Inspired by methodology in hydrodynamical galaxy formation simulations \citep[e.g.][]{Einasto84,Springel2001}, we merge components into groups using a friends of friends algorithm. Groups are identified by merging individual galaxies within a certain distance (linking length) to another object. We combine all the flux of the components within such a group and include it as a single object (`system') in our analyses. The detailed investigation of the resolved properties of the clumps within these systems is deferred to a future paper with the full EIGER data.

\begin{figure}
    \centering
    \includegraphics[width=8.3cm]{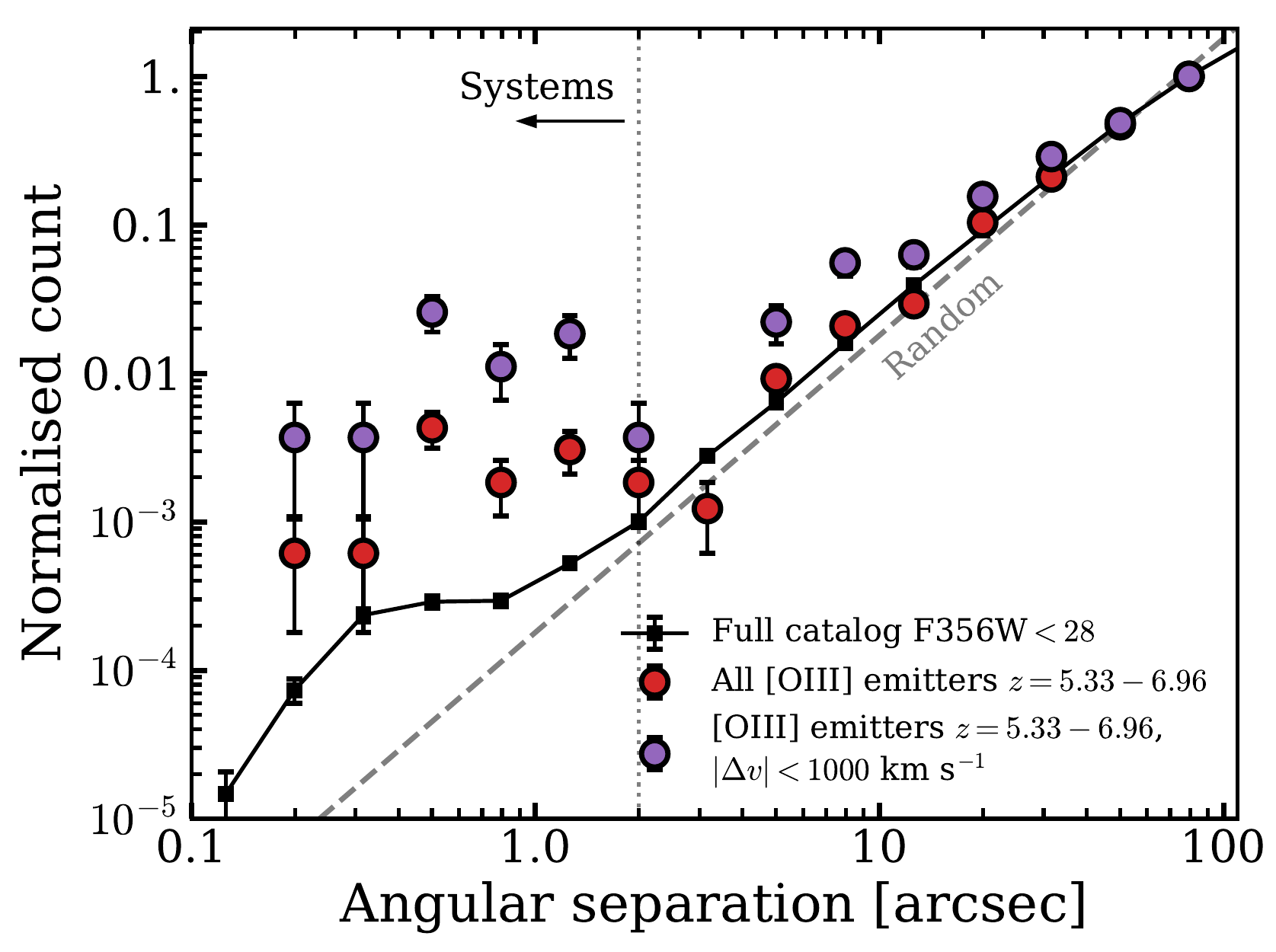}
    \caption{The cumulative distribution of the number of object pairs as a function of their angular separation, normalised to the maximum number of pairs. The red symbols show the distribution of separations for all [OIII] emitters, the purple symbols show the separations for pairs of [OIII] emitters that have velocity differences less than $1000$ km s$^{-1}$ and the black symbols for the full source catalog. Error bars represent poisson noise. The grey dashed line shows the expectation for a random distribution. The [OIII] emitters show a significant excess in the number of pairs below a scale of 2$''$, which corresponds to $\approx10$ kpc. All [OIII] pairs within 2$''$ also happen to be closely separated in redshift ($\Delta v = 600$ km s$^{-1}$ at max).  }
    \label{fig:autocorrelation}
\end{figure} 

The crucial parameter in friends of friends algorithm is the linking length, which in our type of data should both be in the projected and redshift direction. We motivate the maximum linking length based on the projected auto-correlation function of the 133 [OIII] emitting clumps shown in Fig. $\ref{fig:autocorrelation}$. We measure the cumulative distribution of the number of object pairs as a function of their angular separation, normalised to the maximum number of pairs. The auto-correlation function of all [OIII] emitters is compared to the auto-correlation function of the [OIII] emitters with relative velocity differences less than 1000 km~s$^{-1}$ and to all objects in the parent catalog, which are close to randomly distributed. The [OIII] emitters show a clear excess over the full catalog at separations $<2''$ (corresponding to $\sim12$ kpc at $z\sim6$, which is close to the virial radius of halos with mass $10^{11}$ M$_{\odot}$ at $z\sim6$) indicative of amplified small scale clustering, for example due to satellites \citep[e.g.][]{Gelli21}. Interestingly, we find that all [OIII] emitters that are within $2''$ from each other, are also within $|\Delta v|<1000$ km s$^{-1}$ from each other and therefore plausibly physically associated. 

Motivated by these results, we match all [OIII] emitters within a linking length of 2$''$ and merge the individually detected components within such groups together. As a result, 27 of the 133 [OIII] emitters are merged to 13 groups, yielding a final sample of 117 [OIII] emitting systems.

\subsection{ Flux Measurements} \label{sec:measurements}
We measure total line-fluxes of the galaxies from the grism EMLINE data using an optimal 1D extraction as follows. 
First, we collapse the profile of the [OIII]$_{5008}$ line over $\pm400$ km s$^{-1}$ in the spectral direction ($\pm 1000$ km s$^{-1}$ for multiple component systems) and use the \texttt{python} package \texttt{lmfit} to fit the spatial profile with between 1-4 gaussians based on visual inspection of the grism and imaging data and the relative goodness of fit. The vast majority of objects are small and unresolved and therefore fitted with a single gaussian. We use the shape of the [OIII]$_{5008}$ line as it is always the brightest line in the spectra of our galaxy sample. Measurements are independently performed for modules A and B, and, in case both are available, we average them. Spatially resolved line-ratios will be explored in a future paper.

Then, we extract the full 1D continuum-filtered (EMLINE) spectrum assuming this same spatial profile. Since we noticed that the uncertainties that are propagated by the pipeline significantly under-estimate the noise level, we re-scale the noise level of the 2D EMLINE spectrum by enforcing that the standard deviation of empty-sky pixels that cover wavelengths $\lambda=3.15-3.95 \mu$m in our spectrum equals the mean noise level at the same wavelength range. Example 1D extracted spectra are shown in Fig. $\ref{fig:example_spectra}$.

\begin{figure*}
    \centering
    \includegraphics[width=14.2cm]{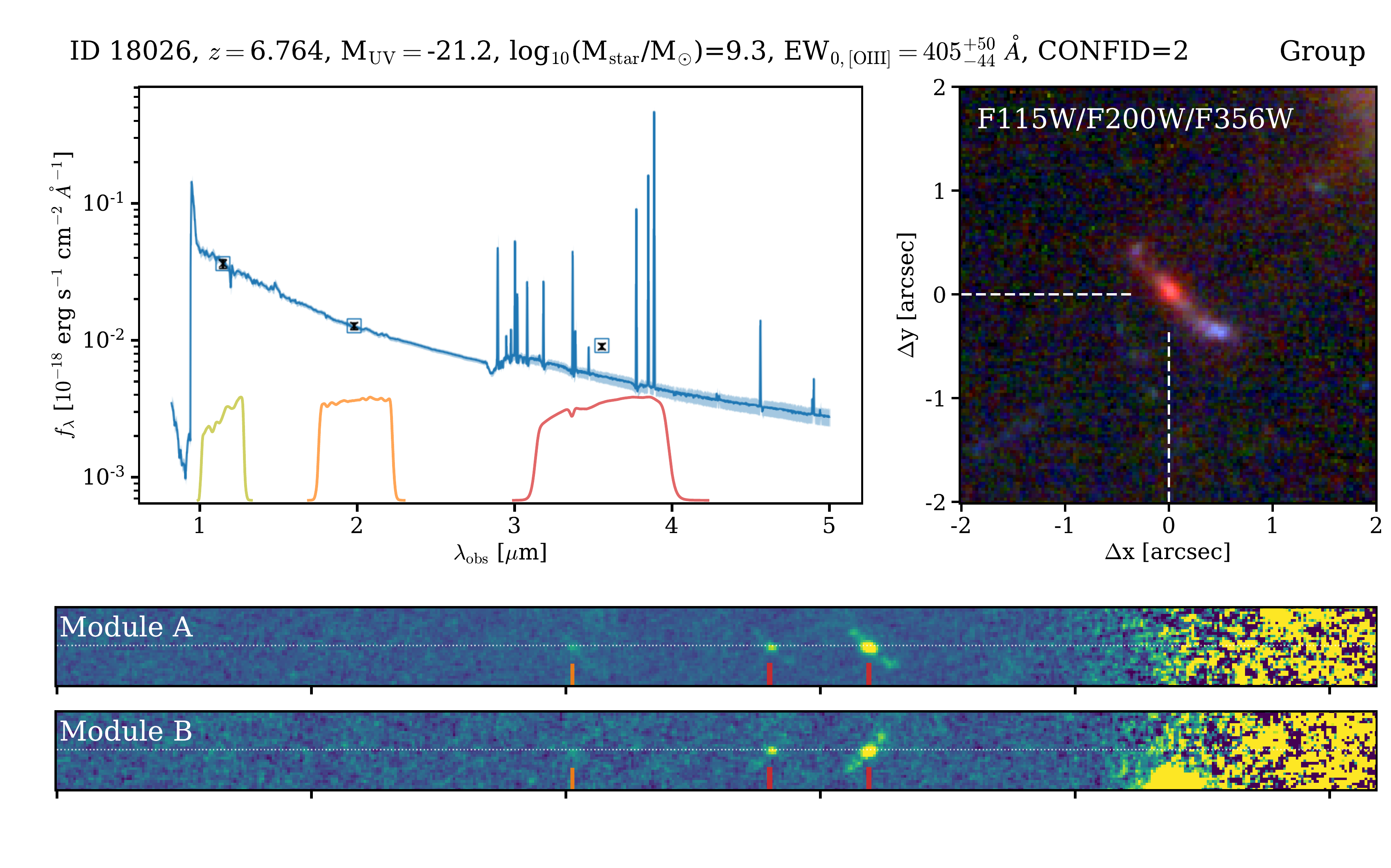}\\\vspace{-0.5cm}
    \includegraphics[width=14.2cm]{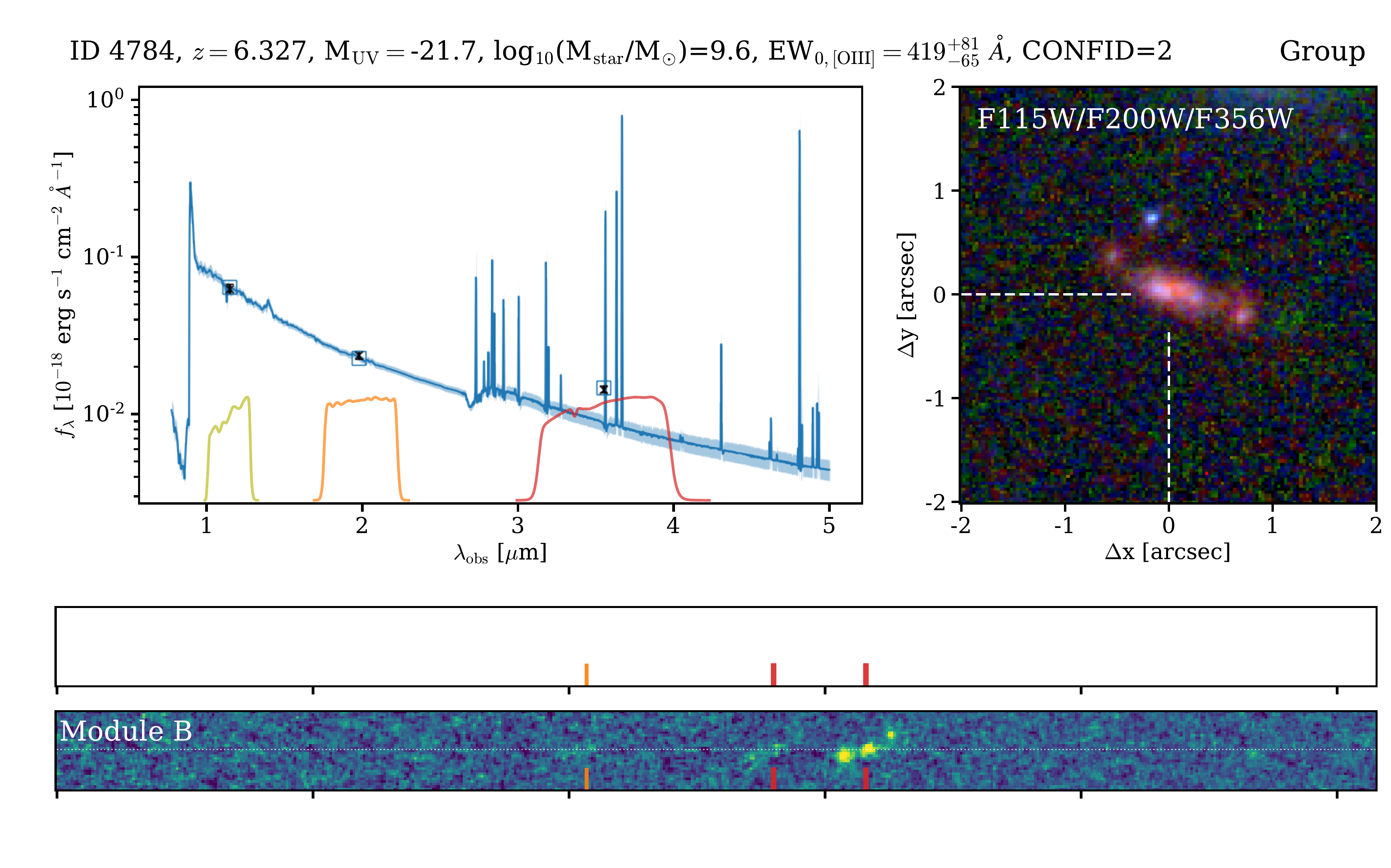} \vspace{-0.3cm}
    \caption{Overview of the information that we can measure for two example [OIII] emitters identified in the {\it JWST} data. For each object, we fit the spectral energy distribution using a composite stellar and nebular emission and dust attenuation model (see \S $\ref{sec:sed}$) to the three photometric data-points and the H$\beta$ and [OIII] emission-line fluxes. Open squares show the modeled flux in the F115W, F200W and F356W filters, respectively, while black squares show the measured photometry and its uncertainties. The main parameters that we derive from the SED models are the UV luminosity, the stellar mass and the emission-line EWs. The false-color stamps reveal the diverse morphologies of the [OIII] emitters whereas the generally red colors highlight the regions within the systems with strong line-emission. We display the 2D continuum-filtered spectra in both modules A and B (when available). The modules have opposite dispersion directions, where module B mirrors the image in the spatial direction. This is clearly illustrated in the spectrum of the object on top. Red and orange lines highlight the locations of H$\beta$ and [OIII], respectively. Both objects in this Figure are identified as groups with three line-emitting components.  }
    \label{fig:SED}
\end{figure*}

\begin{figure*}
    \centering
    \includegraphics[width=14.2cm]{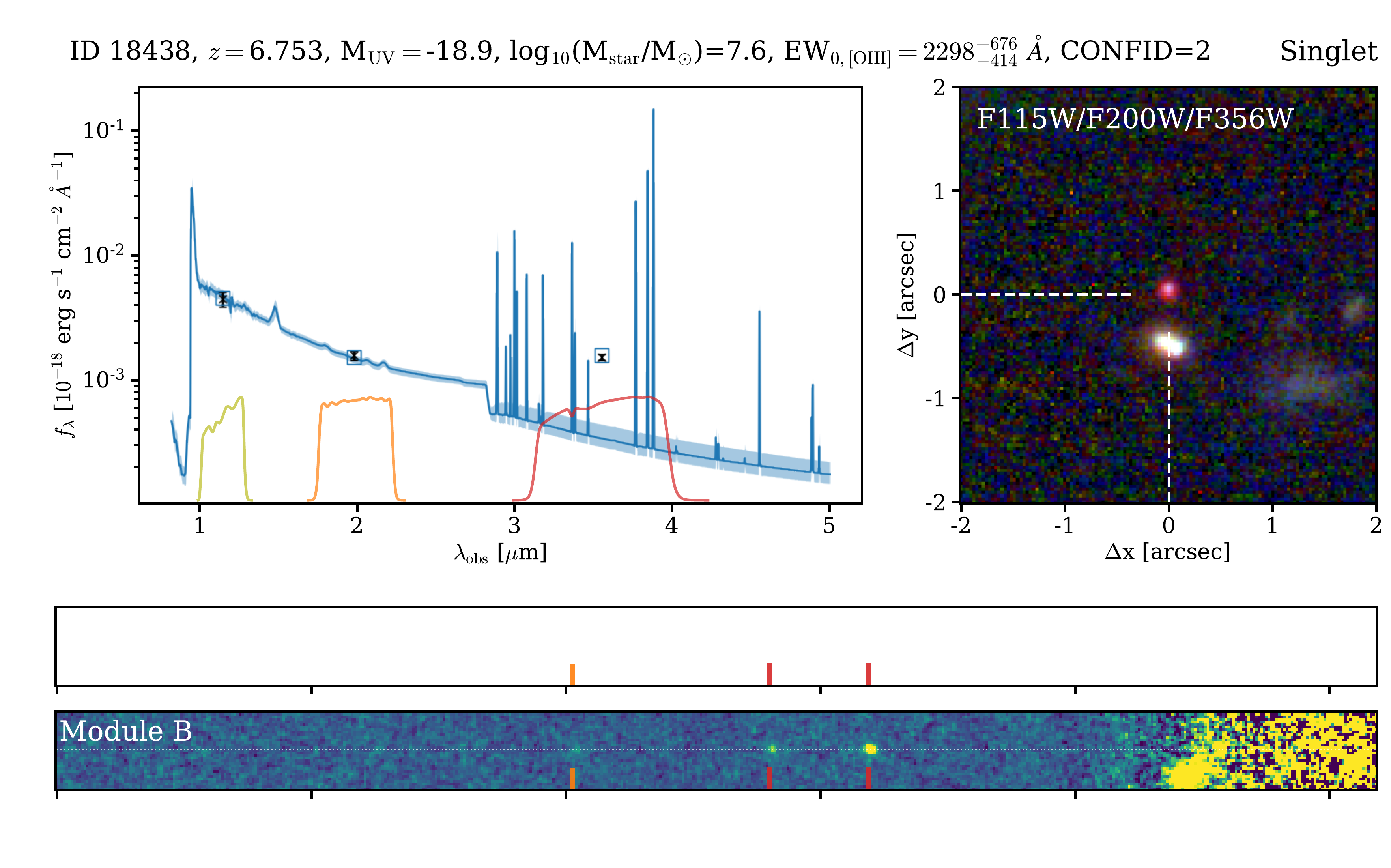}\\ \vspace{-0.5cm}
    \includegraphics[width=14.2cm]{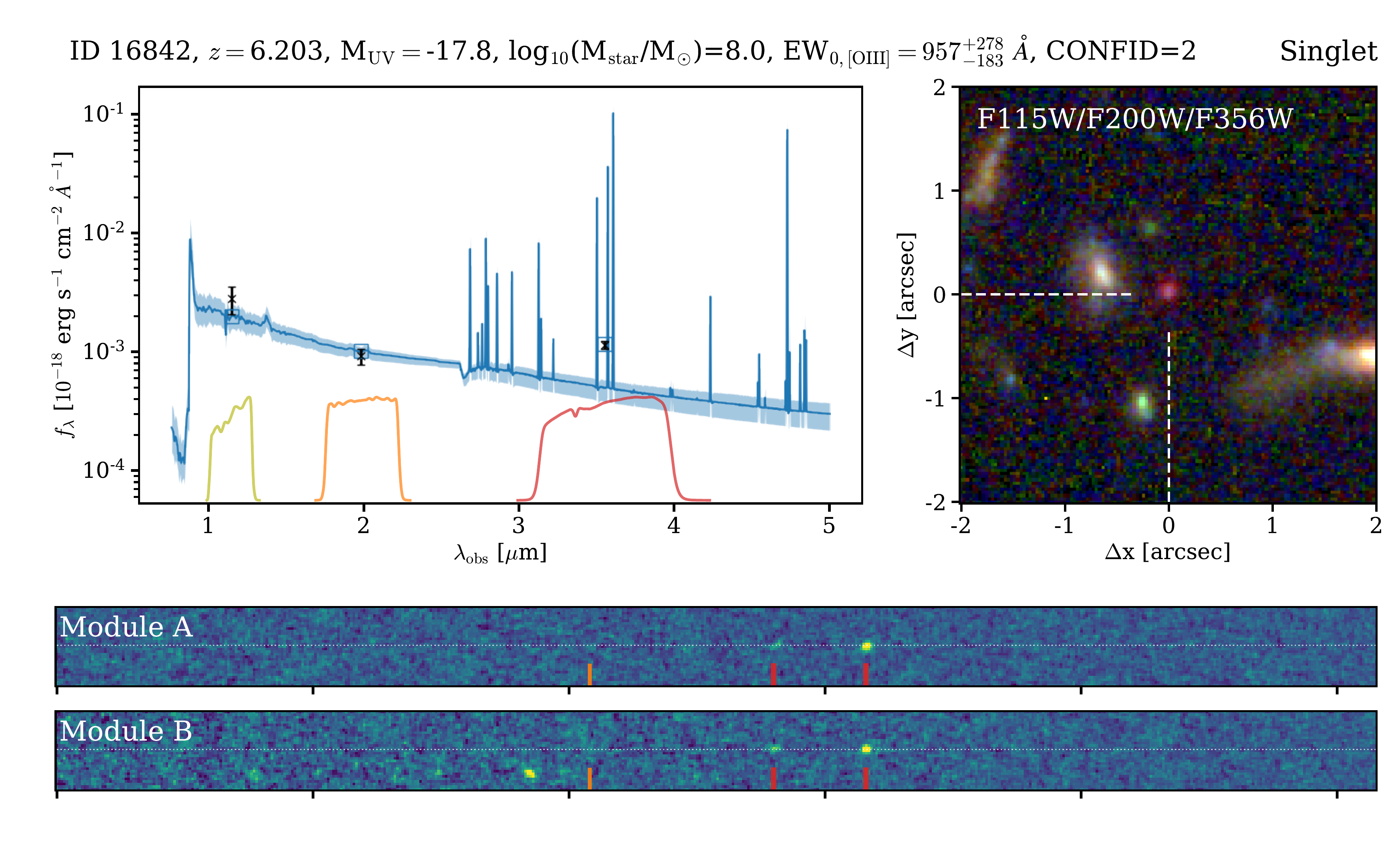}  \vspace{-0.3cm}
    \caption{As Fig. $\ref{fig:SED}$, but now highlighting isolated systems. These sources are representative for the fainter objects in our sample, which are typically small, compact systems with high EWs. }
    \label{fig:SED2}
\end{figure*}

Finally, we fit the spectral line-profile of the [OIII]$_{5008}$ line with between 1-3 gaussians. The fitting is similarly performed using a least-squares algorithm using \texttt{lmfit} and errors on the total line-fluxes are propagated from the covariance matrix. The various components represent multiple closely separated clumps in the grism data, and we find no strong indication for separate dynamical components (e.g. broad wings) in the spectra of individual [OIII] emitters. We then force the same spectral profile on the H$\beta$ and [OIII]$_{4960}$ lines, conservatively allowing 100 km s$^{-1}$ velocity offsets due to possible uncertainties in the wavelength solution, and measure their total line-flux. 

We find that the [OIII]$_{4960}$:[OIII]$_{5008}$ ratio is consistent with the expected 1:2.98 within $1\sigma$ for the vast majority of galaxies (100/117 total objects, where the line-ratio is $2.97\pm0.04$ on average). Uncertainties in the sensitivity curve at the edges of our wavelength coverage, weak contamination by foreground emission-lines or residuals from the continuum filtering process leads to slightly different line ratios in the other 17 objects, with a slight skew towards lower ratios (with a ratio of $1.98\pm0.18$ on average for these 17, $2.87\pm0.07$ for the full sample).

\subsection{SED fitting} \label{sec:sed}
Our survey was designed to obtain a complete sample of spectroscopic redshifts and has less imaging coverage than typical extra-galactic survey fields. For $z\sim6$ galaxies, we cover the rest-frame UV with two filters (F115W, F200W)\footnote{While there is partial coverage in several {\it HST} and ground-based imaging filters, we here only include {\it JWST} imaging data as this leads to the most uniform coverage across the field. At the redshifts considered in this paper, optical data mostly probe wavelengths below Lyman-$\alpha$ and therefore do not help constrain stellar SEDs.} and have one rest-frame optical filter (F356W) that includes the lines covered by our WFSS. This challenges the characterisation of the full spectral energy distributions (SEDs) of our galaxies.

On the other hand, our spectroscopic measurements of nebular line-emission offer significant constraining power on the presence of young stellar populations \citep[e.g.][]{Leitherer1999,Matthee22}. Besides strong emission-lines, the ionizing radiation from the young stellar populations also powers strong nebular continuum emission that may contribute substantially to the spectrum \citep{Reines10,Byler17,Topping22}. For our data, this is particularly relevant because the nebular free-free continuum can boost flux at $\lambda_0\sim3000$ {\AA}, which translates to our F200W photometry. For these reasons, we use a self-consistent inclusion of photo-ionization modeling while fitting SEDs \cite[see also][]{Carnall22,Tacchella22b} using the \texttt{Prospector} code \citep{Johnson21} with nebular treatment based on \texttt{Cloudy} version 13.03 (\citealt{Ferland98}; see \citealt{Byler17} for details).

Our sample selection criteria do not a priori distinguish whether the main dominant source of ionization is star formation or AGN activity. Generally, it is challenging to unambiguously separate AGN from star-burst activity in this context \citep[e.g.][]{Tang22}, since our grism data lacks spectral coverage of important lines as [OII] and H$\alpha$/[NII]. As all lines that we detected have a full width half maximum narrower than $\leq400$ km s$^{-1}$, have low stellar masses and no bright source has a simple point-source morphology, we assume that all galaxies are powered by star formation. 

Specifically, we use \texttt{Prospector} to fit the F115W, F200W and F356W photometry and H$\beta$ and [OIII]$_{4960,5008}$ fluxes. We conservatively estimate the relative grism-to-imaging spectro-photometric calibration to be uncertain at the 5\% level, and increase all errors accordingly. The free parameters in our modeling are the total formed stellar mass, its metallicity and a star formation history, the dust attenuation, the gas-phase metallicity and the ionization parameter. The redshift is fixed to the spectroscopic measurement and we assume a \cite{Chabrier03} initial mass function and MIST isochrones \citep{MIST1,MIST2}. The star formation history follows a delayed-$\tau$ model, i.e. $\psi(t)=\psi_0 t e^{-t/\tau_T}$. The combined nebular and stellar emission are attenuated through a simple dust screen following a \cite{Calzetti00} law. 

We generally use uniform and wide priors: the stellar mass can vary between $10^{6-10.5}$ M$_{\odot}$, the stellar metallicity between [Z/H]=-2.0 and +0.2, the dust optical depth $\tau=0-2$, the age varies between 1 Myr and the age of the Universe at the redshift of each source and the star formation scale factor $\tau_T$ can vary from 100 Myr to 20 Gyr. The gas-phase metallicity spans 12+log(O/H)=6.7 - 9.2 and the ionization parameter $U$ is fit in the range $-3$ to $+1$ in log scale. We allow for these very high values of the ionization parameter to maximise the flexibility of the fits. We finally add a nuisance parameter with values between 0 and 1 that scales the nebular emission relative to that produced by the stellar populations following the \texttt{Cloudy} modeling. This nuisance parameter adds flexibility to account for a non-zero escape of ionizing photons or photoelectric absorption from ionising photons within HII regions \citep[e.g.][]{Tacchella22b}. For slit-based spectra this factor also accounts for the possibility of differential slit loss between continuum and line emitting regions. Our WFSS observations are not affected by slit loss, and therefore provide a useful point of comparison to observations of galaxies at similar redshift made with JWST's {\em NIRSpec} micro-shutter array.

Example SED fits are shown in Figs. $\ref{fig:SED}$ and $\ref{fig:SED2}$. We find that our fits yield SEDs that are characterised by relatively young ages, $t_{\rm age}=110^{+230}_{-80}$ Myr, where the errors show the 16-84th percentiles. The age distribution is typically very narrow ($\tau_T=10.1\pm0.5$ Gyr). Because $t_{\rm age}/\tau_t << 1$ in nearly all cases, the delayed-tau SFR history is well-approximated as a single burst with linearly increasing $\psi(t)$. There is little dust attenuation, E$(B-V)=0.14$ on average. The UV luminosities of our galaxies range from M$_{\rm UV} - 17.7$ to $=-22.3$ (typically M$_{\rm UV}=-19.6$) and the masses span three orders of magnitude from log$_{10}$(M$_{\star}$/M$_{\odot}$)=6.8-10.1 with a median mass of $2\times10^8$ M$_{\odot}$. The nuisance parameter is typically $0.89^{+0.05}_{-0.20}$. The \texttt{Cloudy} models suggest that the gas metallicity and ionization parameter are 12+log(O/H)=7.9 and log(U)=-0.4, but these are both not tightly constrained (i.e. uncertain by almost an order of magnitude in individual objects). This is likely due to complicated dependence of the [OIII]/H$\beta$ line-ratio on metallicity and ionisation parameter. This is further discussed in \S $\ref{sec:metal}$. Additional emission-line measurements not covered by our grism data, such as [OII], are required to better constrain the ionization parameter.

\begin{figure*}
    \centering
    \includegraphics[width=16.3cm]{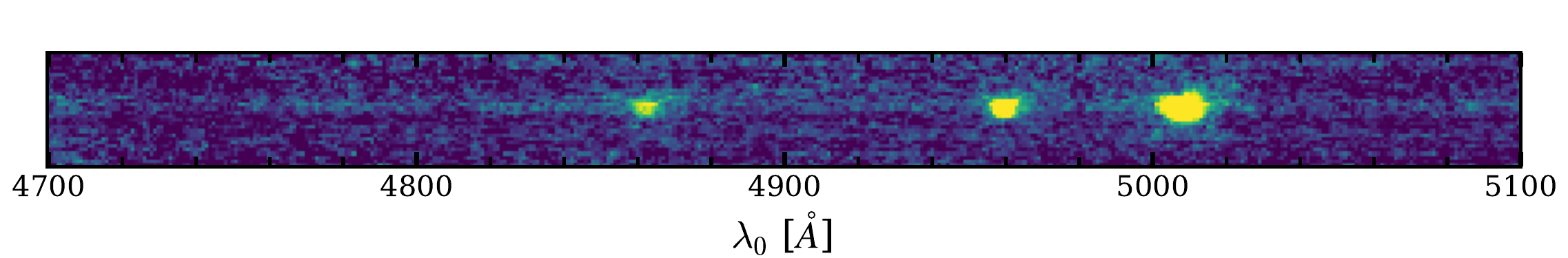}
    \caption{The median stacked spectrum (without continuum filtering) of the 16 [OIII] emitting galaxies at $z=6.28-6.81$ ($z=6.325$ on average) that are not contaminated by foreground objects and have a UV continuum brighter than M$_{\rm UV}<-20.5$. H$\beta$ and the [OIII]$_{4960,5008}$ lines are clearly seen. Continuum emission is detected with a S/N of 5.9 which allows direct spectroscopic measurement of the combined EW$_{0}$(H$\beta$+[OIII])=$948^{+192}_{-138}$ {\AA}.}
    \label{fig:specEW}
\end{figure*}

\section{Strength of Emission lines in $z\sim6$ galaxies} \label{sec:stronglines}
While the presence of strong emission-lines was indicated in broad-band photometry, high-redshift analogues \citep[e.g.][]{Izotov21analog} and the first JWST commissioning spectra \citep[e.g.][]{Sun22b,Tacchella22b,Trump22}, our large spectroscopically confirmed sample finally allows us to investigate the presence of strong rest-frame optical emission-lines in a broad sample of high-redshift galaxies. In this section we first focus on the H$\beta$ and [OIII] EWs, we then present the correlation between UV and [OIII] luminosity and finally present our measurement of the [OIII] luminosity function.

\subsection{Optical line EWs} \label{sec:EW}
\subsubsection{A full spectroscopic EW} \label{sec:specEW}
Typically, SED modeling has been used to estimate the EWs of rest-frame optical lines in high-redshift galaxies \citep[e.g.][]{debarros19}. However, in the absence of line-free photometry redward of the Balmer break, these EWs are uncertain. Hybrid methods, where line-fluxes are measured from spectroscopy while photometry is used to constrain the continuum (\S $\ref{sec:specphotEW}$), are further subject to uncertainties in the relative flux calibrations and aperture effects.

Preferably, the measurement of a line EW is based on spectroscopic data alone as this does not rely on the observatory's (evolving) flux calibration model, but detecting the continuum is challenging when EWs are high. Grism spectroscopy contains significant continuum light from foreground galaxies, further complicating this measurement, as well as the measurement of the actual background. To overcome these limitations, we have visually inspected the WFSS data of all [OIII] emitters and selected the 76 objects that are not strongly contaminated by foreground objects. We have verified that these 76 objects are representative of the full sample in terms of their UV and line luminosities (i.e. their L$_{\rm [OIII]}$-L$_{\rm UV}$ relation is similar to that of the total sample). We then extract 2D SCI spectra of these objects and create {\it median} stacks after masking remaining foreground emission, shifting the spectra to the same rest-frame grid and scaling them by their luminosity distance. A further small background subtraction is applied based on the median value in the off-center part of the stacked 2D spectrum. A median stack is preferred over a mean stack in order to further reduce the impact of unmasked contamination. We then perform an optimally weighted extraction by measuring the average spatial shape of the two [OIII] lines. While we do not detect the continuum in the median stack of all these 76 objects (EW$_0$(H$\beta$+[OIII])$>970$ {\AA} at $5\sigma$ significance), we do detect the continuum for the subset of 16 galaxies with UV luminosity brighter than M$_{\rm UV}<-20.5$, see Fig. $\ref{fig:specEW}$, with a signal to noise of 5.9. The H$\beta$ EW$_0$ of this stack is $116^{+26}_{-19}$ {\AA} and the [OIII]$_{4960+5008}$ EW$_0=832^{+170}_{-122}$ {\AA}. This confirms the extremely high average rest-frame optical emission lines in relatively UV bright galaxies at $z\sim6$ purely using spectroscopy.

\begin{figure}
    \centering
    \includegraphics[width=8.3cm]{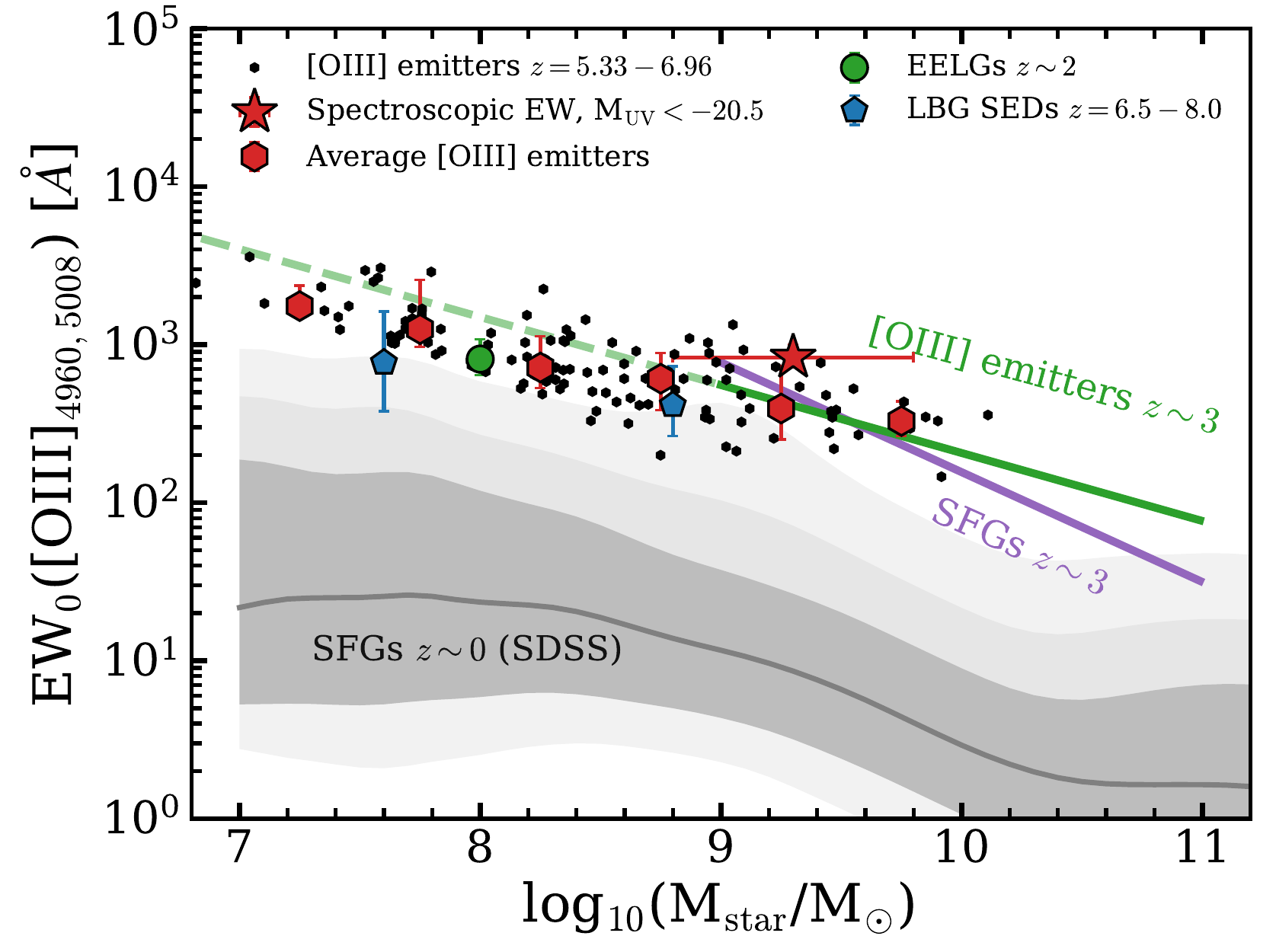}
    \caption{The relation between the [OIII] EW and stellar mass at a wide range of redshifts. The spectroscopically determined EW is shown as a red star. We compare the average EWs in bins of mass (red hexagons; derived as detailed in \S $\ref{sec:specphotEW}$ and error-bars marking the 16-84th percentiles) to extreme emission-line galaxies at $z\approx2$ \citep{vanderWel11} and UV-selected galaxies at $z\sim7$ \citep{Endsley22b}. We show the distribution of EWs in SDSS galaxies at fixed mass (corresponding to the 1, 2, 3$\sigma$ percentiles) in grey shades and also show the mass-dependency of the EW in star-forming galaxies and [OIII]-selected samples at $z\approx3$ \citep[respectively][]{Khostovan2016,Reddy18}.}
    \label{fig:EWmass}
\end{figure}

\subsubsection{Spectro-photometric EWs} \label{sec:specphotEW}
With stacking we lose information on the distribution of EWs and we were further not able to directly measure the EW in the majority of faint galaxies in our sample due to the non-detection of the continuum in the spectrum. We therefore explore EW measurements based on the imaging and WFSS data combined. Line-flux measurements of the lines that contaminate the F356W photometry may directly inform us about the continuum level at these wavelengths and therefore the EW. However, there are uncertainties on the continuum slope within the filter and the contribution from faint undetected lines. We therefore primarily use physically motivated continuum levels based on the SED models (\S $\ref{sec:sed}$) to measure the EWs, but also compare these to more ad hoc measurements.

Our ad hoc measurement of the EW is done as follows: we model the flux measured in the F356W imaging data as a combination of continuum emission that follows a power-law with slope $\beta=-2$ and line emission from H$\gamma$, H$\beta$ and [OIII] and fit this model to the photometry and line-flux measurements. We assume H$\gamma$/H$\beta=0.4$ based on our stacks (\S $\ref{sec:physcond}$). The H$\beta$ and [OIII] EWs are derived from the model output.  Our SED model-based EWs are derived from the posteriors of the SED fitting.

The median SED-based EW of our full sample is EW$_{0}$(H$\beta$+[OIII])=$850^{+750}_{-400}$ {\AA}, which agrees very well with the ad hoc measurement of $840^{+840}_{-440}$ {\AA} (errors correspond to the 68 \% percentiles). The two measurements are typically consistent with a median ratio of 0.98 and a scatter of 0.2 dex. These EWs imply that $40^{+22}_{-15}$ \% of the flux in the F356W photometry is due to line-emission. For the bright subset of 16 galaxies with directly constrained EW, we find that the SED modeling yields EW$_{0}$(H$\beta$+[OIII])=$640^{+390}_{-340}$ {\AA} which is a factor $\approx0.7$ lower than  the directly constrained EW$_{0}$(H$\beta$+[OIII])=$948^{+192}_{-138}$ {\AA}), albeit within the uncertainties. This difference is both due to a lower modeled H$\beta$ and [OIII] EW. 

In Fig. $\ref{fig:EWmass}$ we show average EWs in subsets of mass and show broad agreement with the EWs at $z=6.5-8.0$ derived in recent SED fitting approaches including a continuum-free filter beyond the Balmer break \citep{Endsley22b}, and in low-redshift EELGs \citep{vanderWel11}. The SED derived EWs increase with decreasing mass and UV luminosity, see Fig. $\ref{fig:EWmass}$. The mass dependence of the EW roughly follows the extrapolation of the trend observed at $z\approx3$ \citep{Khostovan2016}, i.e. EW $\propto$ log$_{10}$(M$_{\rm star}$)$^{-0.4}$.

In general, Fig. $\ref{fig:EWmass}$ shows that our measurements support the strong evolution towards extreme emission line galaxies with EWs$\gtrsim1000$ {\AA} becoming {\it typical} at $z\approx6$ (this is further discussed in \S $\ref{sec:ubiq}$), while they only represent $<1$ \% of the galaxies in the SDSS. Our results are also in support of an increasing [OIII] EW with decreasing mass, extending the dynamic range probed at $z\approx3$ by two orders of magnitude to masses $\sim10^7$ M$_{\odot}$ at $z\approx6$.

\subsection{L$_{\rm [OIII]}$-L$_{\rm UV}$ relation}
\begin{figure}
    \centering
    \includegraphics[width=8.3cm]{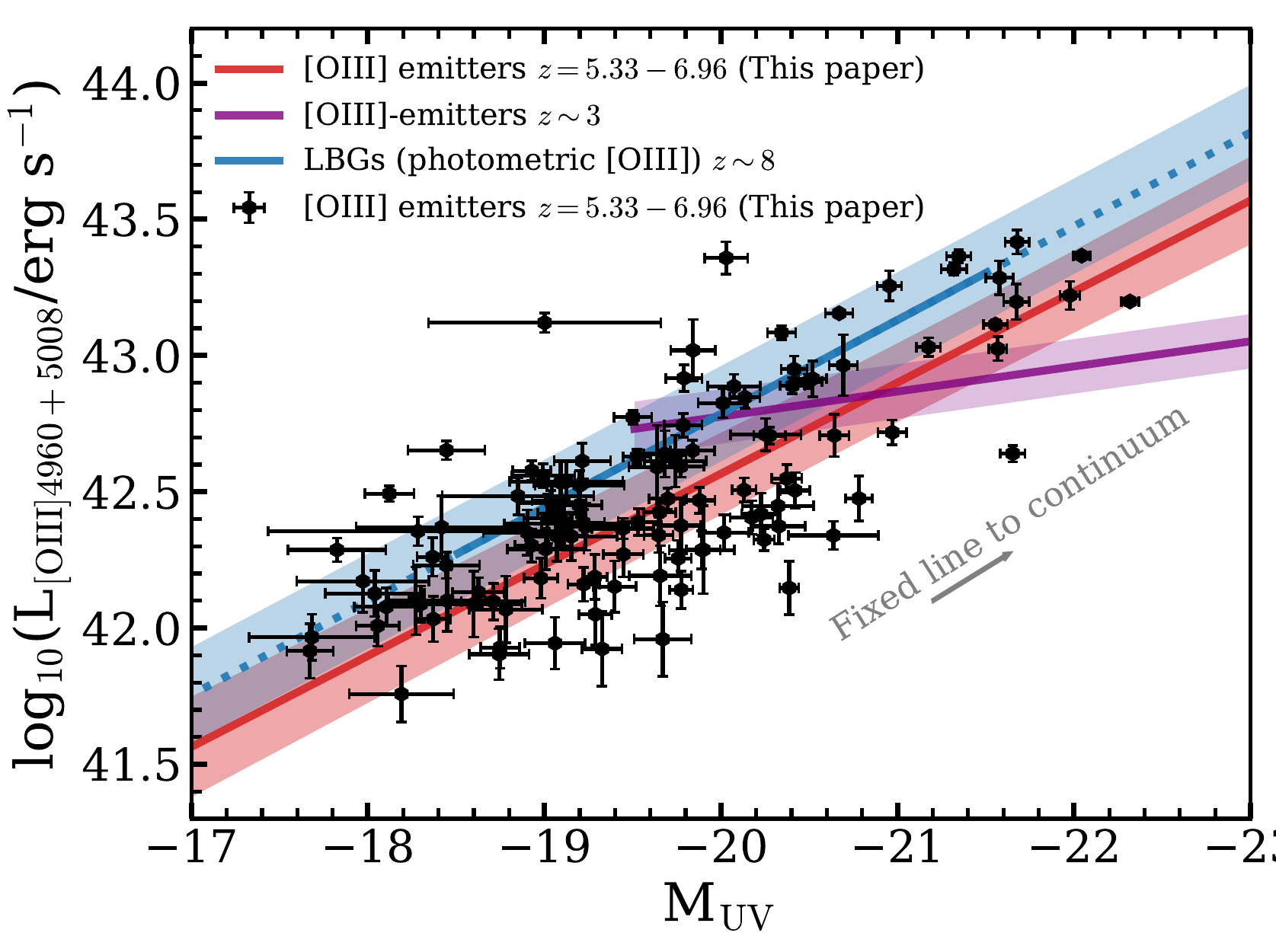}
    \caption{The relation between the [OIII] luminosity and UV luminosity for our sample of [OIII] emitters (black data-points), [OIII] emitters at $z\sim3$ (purple shaded region; \citealt{Khostovan2016}) and the inferred [OIII] luminosities in photometry of UV-selected galaxies at $z\sim8$ (blue shaded region; \citealt{debarros19}). The red line shows a simple linear fit to our data-points log$_{10}$(L$_{\rm [OIII]4960+5008}$/erg s$^{-1}$)$ = 42.60 - 0.30 $(M$_{\rm UV}$+20) with a scatter of 0.27 dex.  }
    \label{fig:MUVLO3}
\end{figure}

Fig. $\ref{fig:MUVLO3}$ compares the measured [OIII] line luminosities with the UV luminosity of the galaxies we identified. Unexpectedly, both luminosities are strongly correlated, roughly following a slope with a fixed [OIII] to UV luminosity ratio. For comparison, Fig. $\ref{fig:MUVLO3}$ also shows the [OIII]-L$_{\rm UV}$ relation for relatively bright [OIII] emitters at $z\sim3$ \citep{Khostovan2016} and the inferred [OIII] luminosities from SED fitting of UV-selected galaxies at $z\sim8$ \citep{debarros19}. Our results point towards a steeper relation compared to $z\sim3$, with a significantly higher [OIII] luminosity at fixed UV luminosity. We interpret this as a likely metallicity effect: at fixed UV luminosity, galaxies at $z\approx3$ plausibly have a higher metallicity which acts to reduce the [OIII] luminosity at a fixed star formation rate. The metallicity of our sample is further investigated in \S $\ref{sec:metal}$.

Our relation between UV and [OIII] luminosity shows a comparable slope to the one measured at $z\sim8$ by \cite{debarros19}, although our spectroscopic measurements are shifted to line-luminosities that are 0.2 dex lower. Their derived EW(H$\beta$+[OIII]) of $\approx650$ {\AA} is comparable to the typical EWs we measure in galaxies with similar UV luminosities (M$_{\rm UV}\approx-20$), suggesting that the differences are actually in the continuum level. Within our sample, the steep slope between [OIII] and UV luminosity implies that there is not a lot of variation in the [OIII] to optical continuum luminosities at $z\sim6$. This could indicate little variation in the [OIII] EWs with UV luminosity, but since the UV-to-optical continuum ratio traces the Balmer break and experiences differences in nebular continuum emission, this trend needs to be interpreted with caution.

\subsection{[OIII] Emission-line Luminosity function} \label{sec:LF}
Here we derive the [OIII] luminosity function at $z\approx6$ and compare it to lower redshifts \citep{Khostovan2016}, earlier estimates at $z\sim8$ \citep{debarros19} and measured number densities \citep[e.g.][]{Sun22b}. 

The luminosity function is measured using the classical 1/V$_{\rm max}$ method \citep{Schmidt68}, where the number density in a given luminosity bin equals
\begin{equation}
    \Phi(L)\, {\rm dlog}L = \sum_i \frac{1}{c_i V_{\rm max, i}},
\end{equation}
where $c_i$ is the completeness of source $i$ and $V_{\rm max, i}$ is the maximum volume in which the source could have been detected. Since our selection criteria relies on the detection of both [OIII] lines, our maximum volume is limited by the redshift range over which both the [OIII]$_{4960,5008}$ lines can be detected ($z=5.33-6.96$). Likewise, a S/N=3 detection of the fainter [OIII]$_{4960}$ line is the limiting factor determining the completeness.

\begin{figure}
    \centering
    \includegraphics[width=8.3cm]{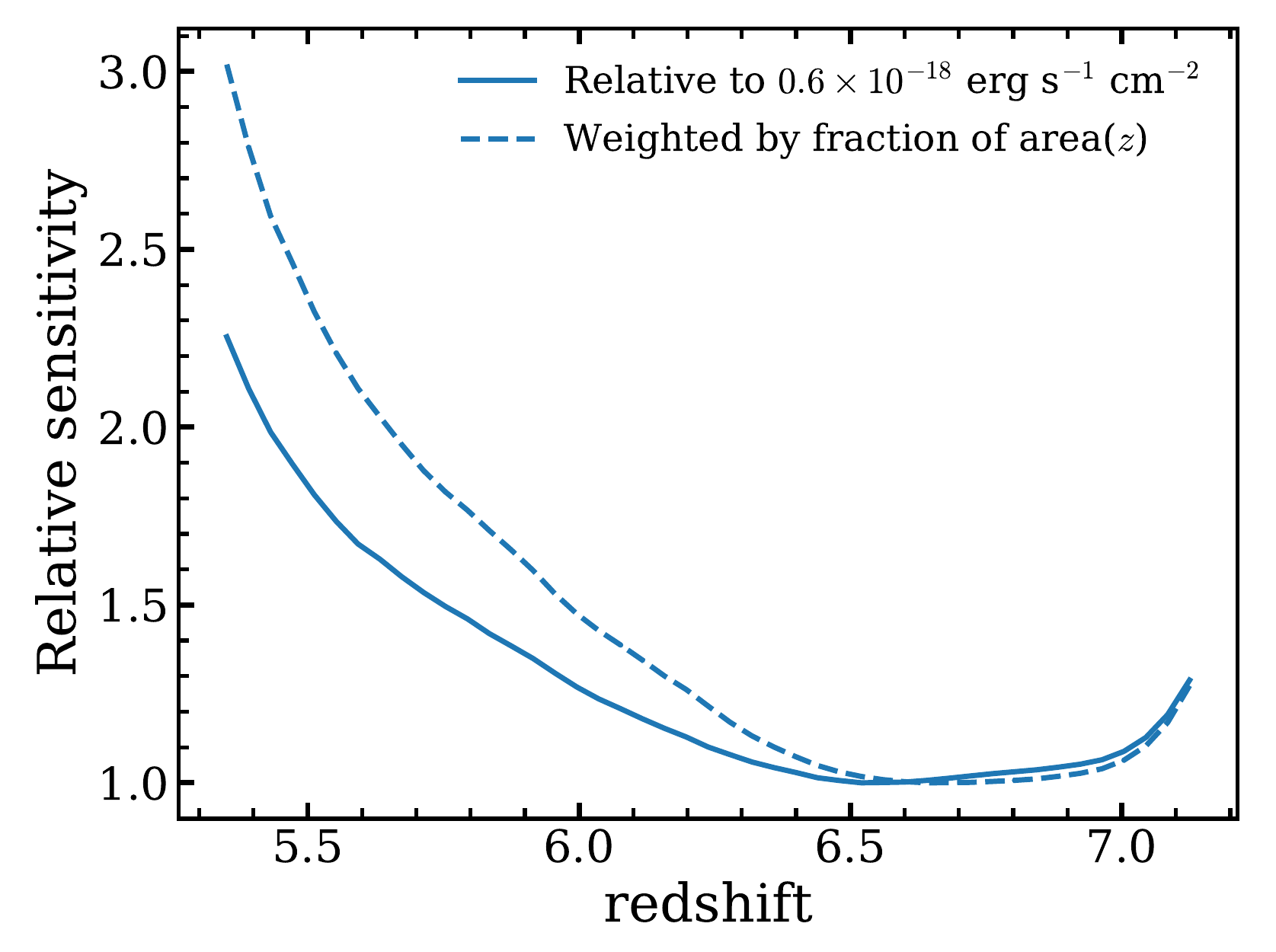}
    \caption{The relative sensitivity of our grism data as a function of wavelength (here converted to the redshift of the [OIII]$_{4960}$ line). The sensitivity is normalised to the maximum sensitivity of a S/N=3 detection for a flux of $0.6\times10^{-18}$ erg s$^{-1}$ cm$^{-2}$ at a redshift of $z\approx6.6$. We illustrate the redshift dependency of the effective field of view of the grism data by weighting the sensitivity with the inverse of the maximum area at each redshift.}
    \label{fig:relsens}
\end{figure}

Our mosaic design yields significant spatial sensitivity variations due to large variations in exposure time and the large $\approx20$ \% difference in the sensitivity of the grism data in module A and module B \citep[e.g.][]{Rigby22}. Furthermore, both the sensitivity and the effective field of view depend on the wavelength of the line. These effects are illustrated in Fig. $\ref{fig:relsens}$, which shows the minimum flux of a S/N=3 detection in our continuum-filtered data as a function of wavelength (here converted to redshift for [OIII]$_{4960}$), relative to the faintest flux detected at that significance ($0.6\times10^{-18}$ erg s$^{-1}$ cm$^{-2}$). The sensitivity is optimal at $z\approx6.4-6.9$ and a factor two lower at $z\approx5.5$. Fig. $\ref{fig:relsens}$ also illustrates the impact of incomplete spectral coverage for a fraction of the field of view, which is increasingly important towards lower redshifts.

\begin{table}[]
    \centering
    \caption{The field number densities of [OIII] emitters at $z=5.33-6.96$ in the first EIGER data in units cMpc$^{-3}$ dlogL$^{-1}$ as a function of [OIII]$_{5008}$ luminosity in erg s$^{-1}$. We mask the redshift range around the quasar J0100+2802. $N$ is the number of objects in each bin, $\langle c\rangle$ the average completeness, $\langle f\rangle$ is the fraction of the area in which a line could be detected, $\Phi_{\rm obs}$ is the number density assuming the total volume and completeness and $\Phi_{\rm corr}$ is the corrected number density taking the completeness and maximum volume into account. The errors combine the poisson noise with a 25 \% uncertainty on the completeness and volume correction added in quadrature. Logarithms are to base 10.}
    \begin{tabular}{crccrr}
     log(L$_{\rm [OIII]}$)   & $N$ & $\langle c\rangle $ & $\langle f\rangle$& log($\Phi_{\rm obs}$) & log($\Phi_{\rm corr}$) \\ \hline
$42.0\pm0.1$ & 12 & 0.38 & 0.72 & $-3.24$ & $-2.67^{+0.18}_{-0.20}$\\
$42.2\pm0.1$ & 18 & 0.45 & 0.82 & $-3.06$ & $-2.63^{+0.14}_{-0.16}$ \\
$42.4\pm0.1$ & 19 & 0.74 & 0.85 & $-3.04$ & $-2.83^{+0.10}_{-0.12}$\\
$42.6\pm0.1$ & 10 & 0.84 & 0.84 & $-3.31$ & $-3.16^{+0.13}_{-0.17}$ \\
$42.8\pm0.1$ & 8 & 0.95 & 0.90 & $-3.41$ & $-3.34^{+0.13}_{-0.19}$ \\
$43.0\pm0.1$ & 8 & 0.95 & 0.89 & $-3.41$ & $-3.34^{+0.13}_{-0.19}$ \\
$43.2\pm0.1$ & 4 & 0.95 & 0.92 & $-3.71$ & $-3.66^{+0.18}_{-0.30} $\\
$43.4\pm0.1$ & 2 & 0.91 & 0.84 & $-4.01$ & $-3.90 ^{+0.23}_{-0.53}$ \\
\end{tabular}
    \label{tab:LF}
\end{table}

We model $c_i$ and $V_{\rm max, i}$ in our data self-consistently with our emission-line selection algorithm as follows. We measure $V_{\rm max, i}$ by creating a 3D data-cube of the 3$\sigma$ line-flux sensitivity as a function of wavelength and position. The cube contains cells of $6.0'' \times 4.5''$. In each cell, we combine all the emission-line line detections in the 2D spectra of the sources whose position is in each cell and measure the flux of the faintest line detected at S/N=3, in steps of $\Delta \lambda_{\rm obs}=20$ nm. In order to overcome shot-noise, we smooth the wavelength dependence of the limiting sensitivity with a uniform filter of 150 nm. To mitigate the higher shot-noise at the edges of our field of view (due to the lower sensitivity), we also smooth the spatial cells with a kernel that increases with the square of the distance to the center. This implies a maximum effective cell size of $22''\times9''$ at the outer edges of our field of view. After creating this cube, we measure the fraction of the total volume for each [OIII]$_{4960}$ line detection at RA$_i$, DEC$_i$ and $\lambda_{{\rm obs},i}$. The completeness for each object, $c_i$, is determined by measuring the detection fraction of fake sources injected in the 2D spectrum at random positions within 250 {\AA} and 1$''$ of the detection line itself. The flux of the fake sources is varied within the 1$\sigma$ confidence interval of the flux of each object. Sources are modeled as 2D gaussians with a FWHM of 0.13$''$ which is representative for the detected lines.

\begin{figure*}
    \centering
    \includegraphics[width=13cm]{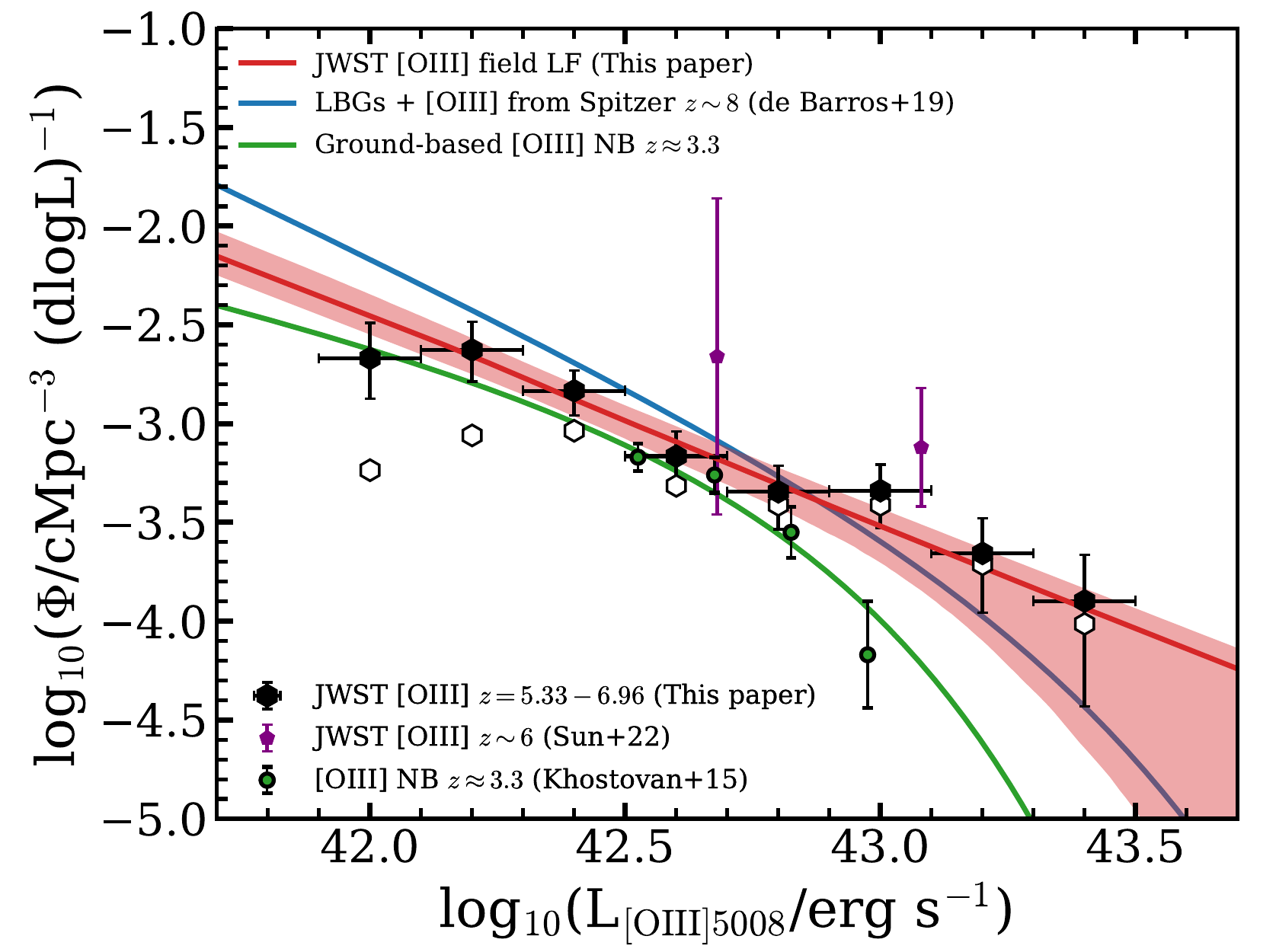}
    \caption{The field [OIII] luminosity function at $z=5.33-6.96$ (masking $\pm1000$ km s$^{-1}$ around the redshift of the quasar; black hexagons). Open hexagons show the uncorrected number densities. The red line and shaded region shows the fitted Schechter function and the 68\% confidence interval fixing the faint-end slope to $\alpha=-2.0$. Our number densities are compared to early {\it JWST} estimates at $z\sim6$ based on commissioning data \citep{Sun22b}, narrow-band measurements at $z\approx3.3$ \citep{Khostovan15} and the inferred [OIII] luminosity function at $z\sim8$ based on modeling the SEDs of UV-selected galaxies \citep{debarros19}. }
    \label{fig:O3_LF}
\end{figure*}

The median completeness of our sample is 79 \% (ranging from 3-99.9 \%) with a primary dependence on line-flux and secondary dependencies on observed wavelength and spatial location. Likewise, the fraction of the total survey volume in which a line can be detected depends mostly on observed wavelength, with a secondary dependence on flux. Typically, this fraction is 84 \% (ranging from 2-97 \%). The maximum volume of our survey spans 25.9 arcmin$^2$ and $z=5.33-6.96$, which corresponds to $1.06\times10^5$ cMpc$^{3}$.

As our survey targets a well known luminous quasar at $z=6.33$, our number densities of the total sample are likely not representative. Indeed, the quasar is embedded in a significant galaxy over-density (which will be discussed in detail in Mackenzie et al. in prep). To obtain an unbiased measurement of the luminosity function, we therefore mask the redshift range $z=6.30-6.35$, i.e. $\pm1000$ km s$^{-1}$ around the quasar and yield a maximum survey volume of $1.03\times10^4$ cMpc$^3$. Further, in order not to depend on uncertainties in our completeness estimate, we only include sources with a completeness of $>10$ \% and remove the few objects identified thanks to H$\beta$ (where S/N [OIII]$_{4960} < 3$). This leaves 84 sources that were used to measure the [OIII] luminosity function. We measure number densities in bins of 0.2 dex and report both the raw number of objects in each bin as well as the raw and corrected number densities in Table $\ref{tab:LF}$.

Typically, luminosity functions are parametrised with a \cite{Schechter76} function:
\begin{equation}
\Phi(L) dL = \Phi^{\star} (\frac{L}{L^{\star}})^{\alpha} e^{-\frac{L}{L^{\star}}} d(\frac{L}{L^{\star}}),
\end{equation}
where $\Phi^{\star}$ is the characteristic number density, $L^{\star}$ the characteristic luminosity and $\alpha$ the faint-end slope. As shown in Fig. $\ref{fig:O3_LF}$, our  measured number densities of [OIII] emitters at $z\sim6$ do not show significant evidence for an exponential decline at high luminosities. Therefore, fitting all three parameters of the LF (which we do in linear space) leads to significant uncertainties ($\alpha=-0.24^{+1.24}_{-1.88}$, log$_{10}(\Phi^{\star})=-2.77^{+0.17}_{-5.27}$ cMpc$^{-3}$ and log$_{10}(L^{\star}$/erg s$^{-1})=42.18^{+3.49}_{-0.33}$). When we fit the LF while fixing the faint-end slope to $\alpha=-2.0$ which is the same as the UV LF at $z\sim6$ \citep[e.g.][]{Bouwens21}, we obtain the following constraints log$_{10}(\Phi^{\star})=-7.74^{+4.01}_{-0.15}$ cMpc$^{-3}$ and log$_{10}(L^{\star}$/erg s$^{-1})=46.94^{+0.05}_{-3.90}$. While the fitted faint-end slope is uncertain, we note that a flatter faint-end slope compared to the UV LF may not be unexpected in case the mass-metallicity relation is steep and lower mass galaxies emit increasingly less [OIII] photons. This is further explored in \S $\ref{sec:metal}$.

When comparing our measured [OIII] LF to results at lower redshifts \citep[e.g. $z\approx3$][]{Khostovan15}, we find remarkably little evolution in the [OIII] LF over $z=3-6$, in stark contrast to the decline of the UV LF and the cosmic star formation rate density over this epoch \citep[e.g.][]{Finkelstein16,Bouwens21}. Our number densities are lower than early {\it JWST} commissioning results \citep{Sun22b} and are comparable to those inferred by \cite{debarros19} based on SED modelling at $z\approx8$, although with a clearly different shape. We discuss the interpretation of the [OIII] luminosity function and the caveats further in \S $\ref{sec:discuss}$.

\begin{figure*}
    \centering
    \includegraphics[width=18cm]{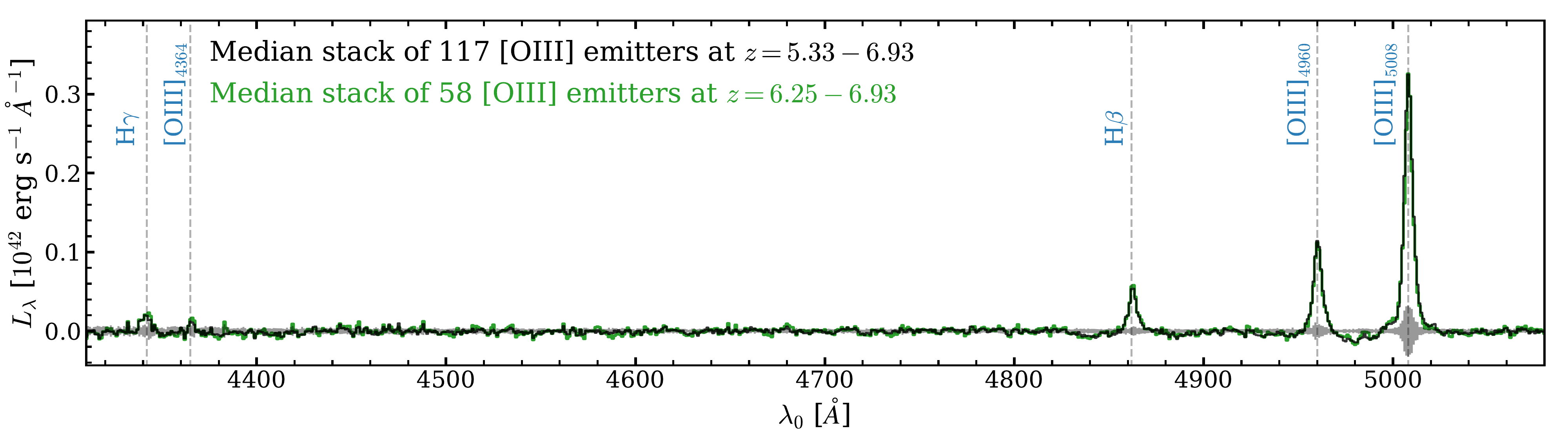}
    \caption{Median stacked 1D rest-frame emission-line spectrum of the full sample of 117 [OIII] emitters at $z=5.33-6.93$ (black) and the subset of 58 galaxies at $z=6.25-6.93$ that has uniform coverage over the observed wavelength range shown here (green). The grey shaded region shows the uncertainty estimated through bootstrap resampling. Each spectrum is weighted equally. We highlight the wavelengths of H${\gamma}$, [OIII]$_{4364}$, H$\beta$ and [OIII]$_{4960,5008}$ which are all detected at an integrated S/N$>5$.} 
    \label{fig:stacked_1D}
\end{figure*}

\section{The physical conditions in early galaxies} \label{sec:physcond}
In addition to selecting and confirming the distances to galaxies, the measured emission-lines also provide insights into the properties of galaxies. In this section we first compare the variations in the H$\beta$-to-UV luminosity ratios and interpret this in the context of the production efficiency of ionising photons. We then focus on line-ratios, in particular [OIII]/H$\beta$, which are interpreted in the context of the gas-phase metallicity. Here we limit ourselves to galaxies at $z>5.5$ where H$\beta$ is spectrally covered.

\subsection{Stacking methodology}
While we base our results on detections of [OIII] and H$\beta$ in many individual galaxies, we also use results from {\it median} stacks in various subsets of our galaxy sample as these stacked spectra allow the detection of fainter features. In Fig. $\ref{fig:stacked_1D}$ we show the stacked 1D spectrum of the full sample of [OIII] emitters. Since the rest-frame wavelength coverage bluewards of [OIII] is not uniform, one needs to interpret this stack with caution, but we show it for illustrative reasons here. The median stack is obtained after shifting the 1D spectrum of each [OIII] emitter to the same rest-frame wavelength grid and rescaling it with the luminosity distance and redshift. We also account for slight residuals in the overall background and continuum subtraction by subtracting the median continuum level masking wavelengths of possible emission lines. Uncertainties are estimated by stacking 1000 bootstrap realisations of the sample. Fig. $\ref{fig:stacked_1D}$ shows the strong H$\beta$ and [OIII]$_{4960,5008}$ emission-lines, but also evidence for H$\gamma$ and [OIII]$_{4364}$ which we will discuss below. We note that faint residuals of our continuum-filtering method are present around H$\beta$ and [OIII]. These are due to our emission-line mask missing some of the faint outskirts and/or companions of the emission-line detections while performing the continuum filtering. We also note that the line-profiles of the stronger lines appear to have faint broad wings, but their interpretation is ambiguous as any spatial extent of the lines in the dispersion direction could mimic such dynamical features. We therefore only focus on line-fluxes, which we measure using a two component gaussian model. By comparing the EMLINE and SCI stacks of the clean sample (see \S $\ref{sec:specEW}$), we find that these residuals affect the flux of the [OIII]$_{5008}$ line by 5 \%, and other lines by $<1$ \%. We therefore base the [OIII] flux measurements of all our stacks on the measured [OIII]$_{4960}$ flux and assume the intrinsic 1:2.98 ratio, verified in the stack of the clean sample. The measurements of the UV luminosity, median stellar mass and H$\beta$ and [OIII] luminosities are listed in Table $\ref{tab:stack1}$, where we also list various properties derived from these measurements.

\begin{figure}
    \centering
    \includegraphics[width=8.5cm]{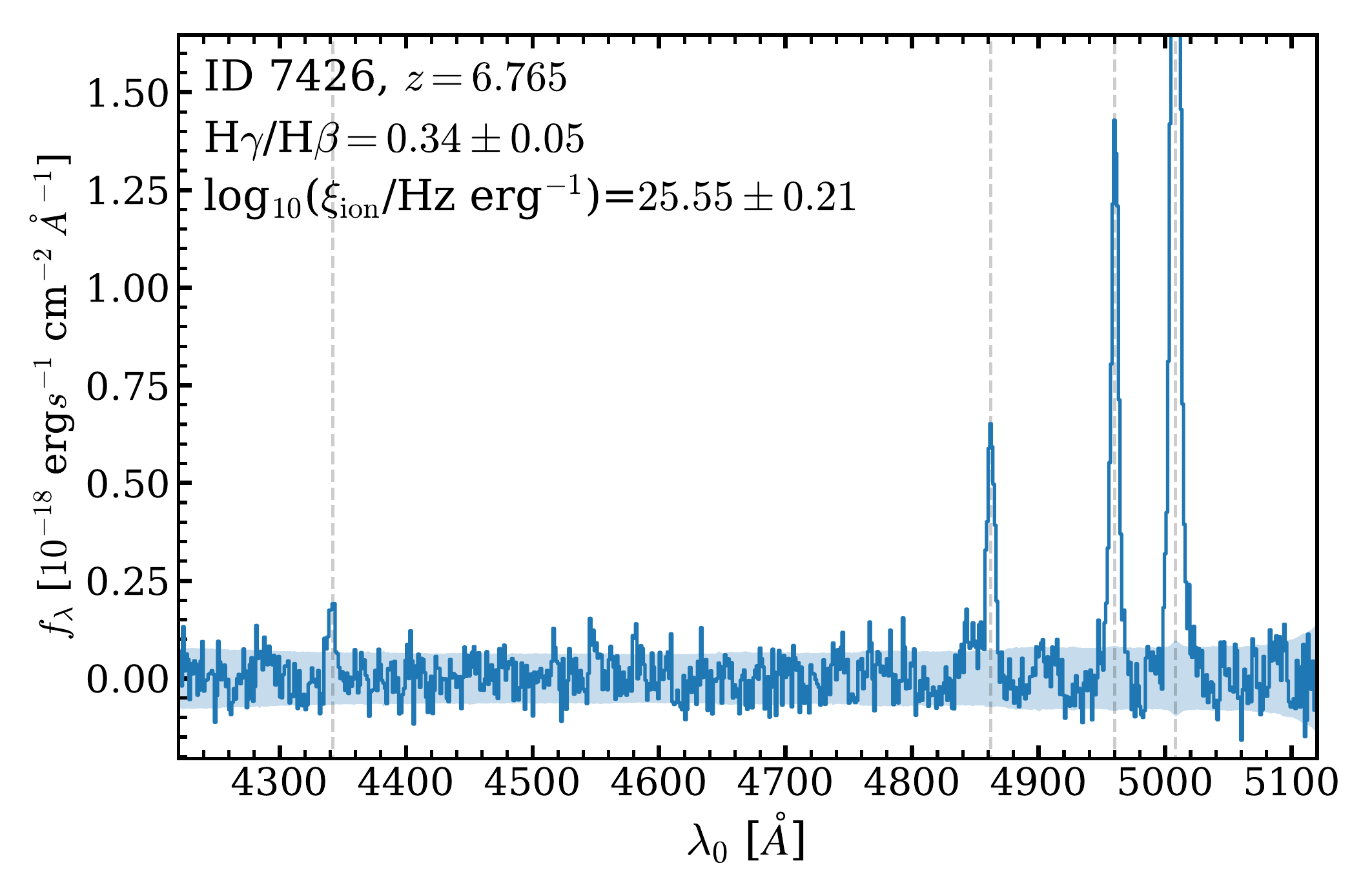}
    \caption{1D spectrum of one of the most luminous [OIII] emitters in our sample which shows detections of H$\gamma$ (integrated S/N=7.6), H$\beta$ and [OIII]$_{4960,5008}$. Note that the region of H$\gamma$ and [OIII]$_{4364}$ is covered only for galaxies at $z>6.25$. } 
    \label{fig:stacked_7426}
\end{figure}

\begin{table}[]
    \centering
    \caption{Measurements of the median stack of our full sample of [OIII] emitters from $z=5.33-6.93$ and the subset at $z>6.25$ for which H$\gamma$ and [OIII]$_{4364}$ are measured. Line luminosities are not corrected for attenuation. For the full sample, the H$\beta$ luminosity is dust-corrected assuming E($B-V)$=0.1 based on the SED fits when measuring SFR and $\xi_{\rm ion}$. As detailed in \S $\ref{sec:xion}$, we use an appropriate conversion between H$\beta$ luminosity and SFR for the typical ionizing photon production efficiency. }  \raggedright
    \begin{tabular}{lrr}
      Property   & Full sample & $z>6.25$  \\ \hline
      N & 117 & 58 \\
     M$_{\rm UV}$     &  $-19.6\pm0.1$ & $-19.5\pm0.1$ \\
     L(H$\gamma$)/$10^{42}$ erg s$^{-1}$ & - & $0.17\pm0.03$ \\
     L([OIII]$_{4364}$)/$10^{42}$ erg s$^{-1}$ & - & $0.08\pm0.01$ \\ 
     L(H$\beta$)/$10^{42}$ erg s$^{-1}$ & $0.38\pm0.02$ & $0.41\pm0.02$ \\ 
     L([OIII]$_{4960,5008}$)/$10^{42}$ erg s$^{-1}$ & $3.13\pm0.16$ & $2.86\pm0.18$ \\
     log$_{10}$([OIII]$_{5008}$/H$\beta$) & $0.80\pm0.04$ & $0.71\pm0.04$  \\    \hline
     log$_{10}$(M$_{\star}$/M$_{\odot}$) & $8.38\pm0.07$ & $8.45\pm0.09$ \\ 
     EW$_{0}$(H$\beta$+[OIII]$_{4960,5008}$)/{\AA} & $845\pm70$ & $860\pm90$ \\
     E$(B-V)_{\rm gas}$ & - & $0.14^{+0.16}_{-0.14}$\\
     SFR(H$\beta$)/M$_{\odot}$ yr$^{-1}$ & $4\pm1$ & $5^{+4}_{-2}$  \\  
     log$_{10}(\xi_{\rm ion}$/Hz erg$^{-1})$ & $25.3\pm0.2$  & $25.31^{+0.29}_{-0.16}$ \\   
     $T(O^{++})$/$10^4$ K & - & $2.2\pm0.2$ \\
12+log$_{10}$(O/H)$_{\rm O3H\beta}$ & $7.41\pm0.10$ & $7.25\pm0.08$ \\
12+log$_{10}$(O/H)$_{\rm direct}^{\ddagger}$ & - & $7.38\pm0.09$ \\ \hline
    \end{tabular}
{\raggedright $\ddagger$ \footnotesize{Based on assuming [OIII]/[OII]$=8\pm3$.} \par}
    \label{tab:stack1}
\end{table}

\subsection{The production efficiency of ionising photons} \label{sec:xion}
Our SED fitting results indicate that our sample of [OIII] emitters is characterised by relatively young stellar ages $\sim100$ Myr that power the strong emission-lines. Here we interpret the emission-line strengths in the context of the production efficiency of ionising photons, $\xi_{\rm ion}$, which is a crucial parameter in assessing the impact of galaxies on cosmic reionization \citep[e.g.][]{Robertson13}. $\xi_{\rm ion}$ can be measured from stellar population models \citep[e.g.][]{duncan&conselice15}, but also using the Balmer recombination lines \citep[e.g.][]{Bouwens16b}. 

We detect the H$\beta$ line in a significant fraction of the objects in the sample and can therefore estimate $\xi_{\rm ion} = \frac{L_{\rm H\beta}}{c_{\rm H\beta} L_{\rm UV}}$, where the line-emission coefficient $c_{\rm H\beta} = 4.86\times10^{-13}$ erg for case B recombination with electron temperature $10^4$ K and a zero escape fraction of ionising photons \citep[e.g.][]{Schaerer03}. Without applying dust corrections, we measure an average log$_{10}(\xi_{\rm ion}$/Hz erg$^{-1})=25.5\pm0.1$.

\begin{figure*}
    \centering
    \includegraphics[width=13cm]{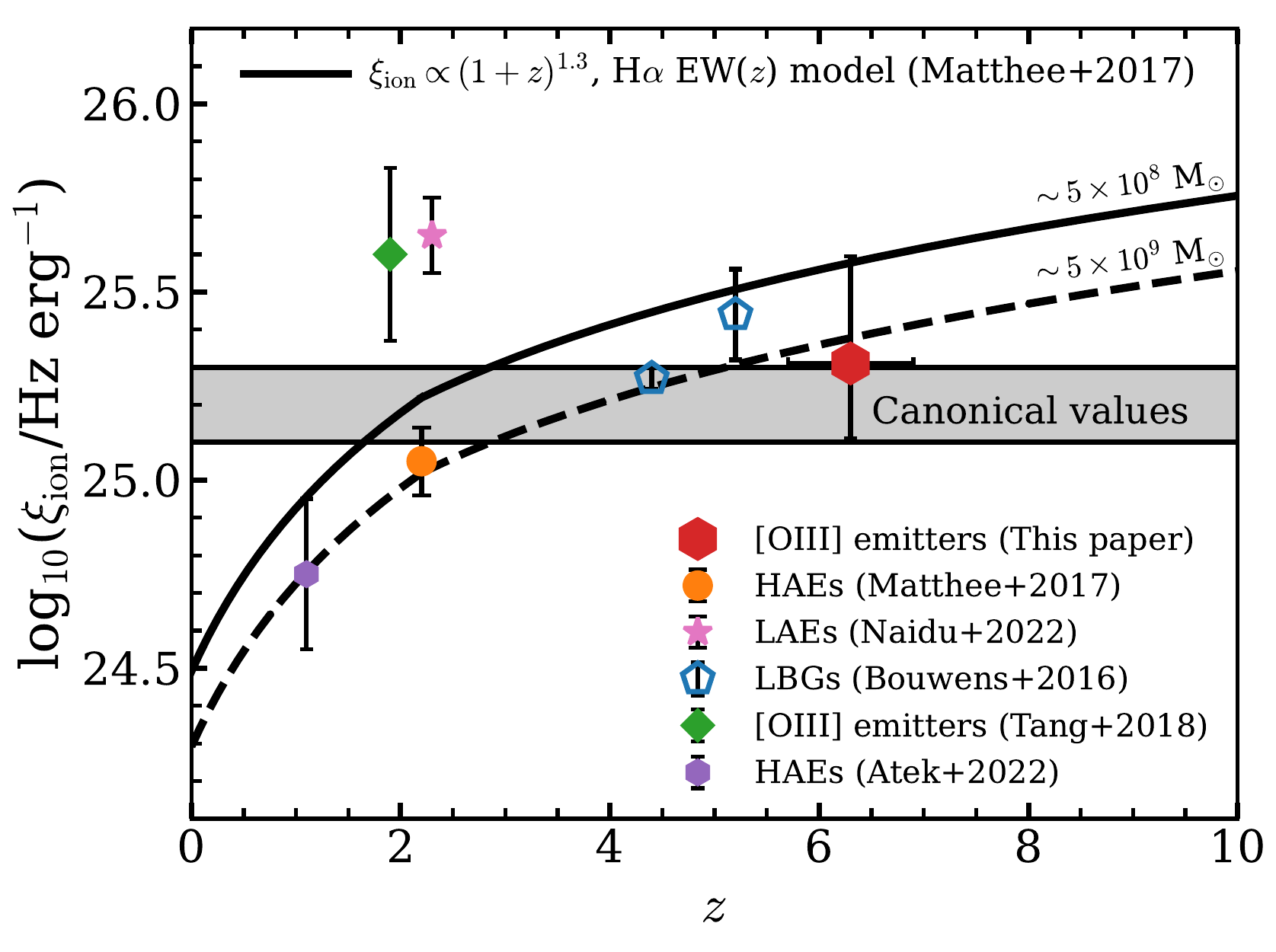}
    \caption{The ionising photon production efficiency $\xi_{\rm ion}$ of various galaxy populations through cosmic times. Spectroscopic measurements are shown as filled symbols, while photometric inferences \citep[e.g.][]{Bouwens16b} are shown as open symbols. The average value of $\xi_{\rm ion}$ measured in our sample of [OIII] emitters is somewhat higher than the canonical value \citep[e.g.][]{Robertson13} and normal galaxies at low-redshift \citep{Matthee17,Atek22a}, comparable to LBGs at $z\approx5$ and somewhat lower than to $\xi_{\rm ion}$ measured in extreme emission-line galaxies at $z\approx2$ \citep{Tang19,NaiduMatthee22}. }
    \label{fig:Xion}
\end{figure*}

The dust correction is however critical in measuring $\xi_{\rm ion}$ accurately \citep{Matthee17,Shivaei18}. The stellar attenuation has been estimated for individual sources from the SED modeling and is typically low, E$(B-V)_{\star}=0.11^{+0.09}_{-0.07}$. Two objects have H$\gamma$ detections that allow us to derive their nebular attenuation using the Balmer decrement and the observed H$\gamma$/H$\beta$ ratio. Assuming an H$\gamma$/H$\beta=0.47$ in the unattenuated case, we measure E$(B-V)=0.30^{+0.13}_{-0.13}$ (Fig. $\ref{fig:stacked_7426}$) and E$(B-V)=0.00^{+0.05}_{-0.00}$ for these two sources, respectively. These attenuations are similar to the stellar attenuations in the SED modeling and imply log$_{10}(\xi_{\rm ion}$/Hz erg$^{-1})=(25.39 - 25.55)\pm0.20$ for these two sources in case the nebular and stellar attenuation follow \cite{Cardelli89} and \cite{Reddy16a} attenuation curves, respectively. 

In order to obtain a sample average, we estimate the typical nebular attenuation based on a stacked spectrum of the subset of 58 [OIII] emitters at $z=6.25-6.93$ for which we have complete spectral coverage of H$\gamma$. H$\gamma$ is detected with a S/N of 6.4 while the flux of H$\beta$ is measured with a S/N of 17 (see Table $\ref{tab:stack1}$). The observed H$\gamma$/H$\beta$ ratio of $0.41\pm0.06$ is, albeit uncertain, suggestive of little attenuation E$(B-V)_{\rm gas}=0.14^{+0.16}_{-0.14}$, consistent with the stellar attenuation. If we thus assume an attenuation of E$(B-V)=0.1$ for the total sample, we find a typical log$_{10}(\xi_{\rm ion}$/Hz erg$^{-1})=25.31^{+0.29}_{-0.16}$. We note that more precise measurements require better constraints on the attenuation.

This measurement of $\xi_{\rm ion}$ constitutes the first spectroscopic confirmation that the ionizing photon production efficiency in early galaxies appears higher than typical in galaxies at $z\approx0-2$ \citep[e.g.][]{Matthee17,Atek22a}, see Fig. $\ref{fig:Xion}$. Measurements of $\xi_{\rm ion}$ in rare high-redshift analogues such as green pea galaxies and Lyman-$\alpha$ emitters at $z\approx0-2$ with high [OIII] EWs that are comparable to the lower mass galaxies in our sample (Table $\ref{tab:MZR}$) show higher values of $\xi_{\rm ion}$ \citep{Schaerer16,Tang19,Matthee21}, suggestive of significant variation among the galaxies. Our results are along the trend of increasing $\xi_{\rm ion}$ at fixed mass \cite{Matthee17} following the evolution of the H$\alpha$ EW.

The ionizing photon production efficiency is also related to the measurement of star formation rate through the Balmer lines. In the standard SFR calibrations from \cite{KennicuttEvans12}, equating the UV SFR to the H$\alpha$ SFR implies log$_{10}(\xi_{\rm ion}$/Hz erg$^{-1})=25.12$. The higher measured $\xi_{\rm ion}$ implies that fewer stars are required to power the same H$\alpha$ (or H$\beta$) luminosity \citep[e.g.][]{Theios2019}. Following the scaling in Table 2 from \cite{Theios2019}, we assume a SFR-L(H$\alpha$) conversion of log$_{10}$(L(H$\alpha$) M$_{\odot}$ yr$^{-1}$/erg s$^{-1}$)=41.59 appropriate for our measured value of $\xi_{\rm ion}$. Computing the H$\alpha$ luminosity from the measured H$\beta$ luminosity and assuming E$(B-V)=0.1$, this implies a typical SFR(H$\beta$)=$4\pm1$ M$_{\odot}$ yr$^{-1}$ for the full sample. For the subset at $z>6.25$ where we directly constrain the nebular attenuation, we find a typical SFR(H$\beta$)=$5^{+4}_{-2}$ M$_{\odot}$ yr$^{-1}$.

\begin{figure*}
    \centering
    \includegraphics[width=13.3cm]{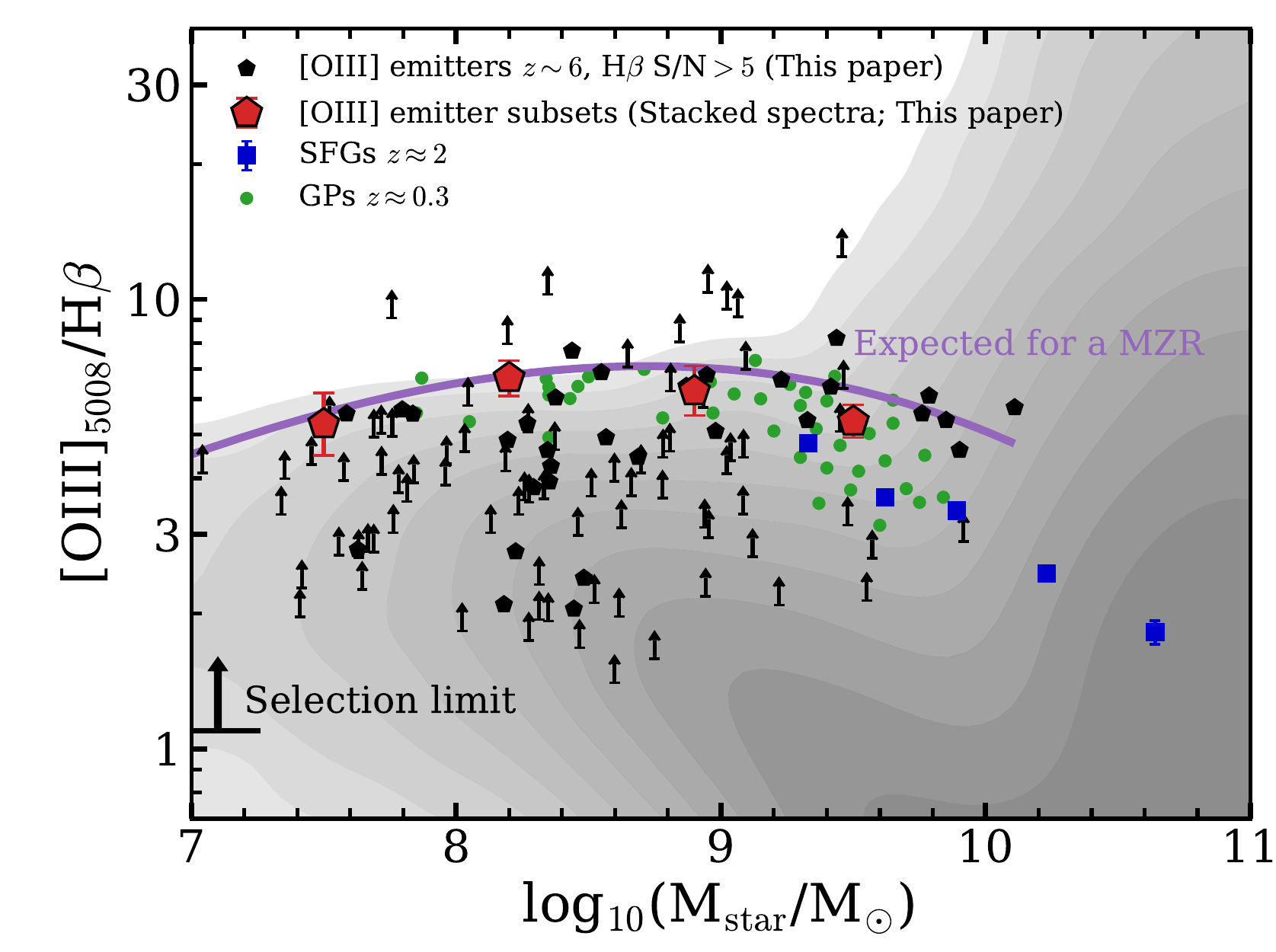}
    \caption{The location of our sample of [OIII] emitters (black hexagons for H$\beta$ detections with S/N$>5$, arrows otherwise) on the Mass-Excitation diagram compared to galaxies in the SDSS (grey shades), green pea galaxies at $z\approx0.3$ \citep{Yang2017} and star-forming galaxies at $z\approx2$ \citep{Sanders20}. Red pentagons show the line-ratios in stacks of subsets in mass. Most objects in our sample have a very high [OIII]/H$\beta$ extending the anti-correlation between [OIII]/H$\beta$ and mass identified at $z\approx2$ and comparable to the line ratios measured in green pea galaxies. The objects with the lowest [OIII]/H$\beta$ values are typically found among the intermediate masses in our sample. The purple line shows the expected relation assuming the \cite{Bian2018} strong-line calibration and the $z\approx6$ mass-metallicity relation (MZR) from the FIRE simulation \citep{Ma16}.   }
    \label{fig:MEX}
\end{figure*}

\subsection{Ionisation conditions and gas-phase metallicity} \label{sec:metal}
In Fig. $\ref{fig:MEX}$ we show the locations of our sample of [OIII] emitters on the so-called `MEx' (Mass-Excitation) diagram \citep{Juneau11}. Our sample of [OIII] emitters has an average [OIII]/H$\beta$=6.3, which is very high compared to typical galaxies in the SDSS with similar mass. We stress that a high [OIII]/H$\beta$ ratio was not a strict selection requirement in our galaxy search, and this result thus reflects a physical property of this sample of galaxies. The average [OIII]/H$\beta$ ratio in our sample is similar to typical values measured in green pea galaxies at $z\approx0.3$ \citep[e.g.][]{Yang2017}. We also show the measurements in the stacked spectra of four subsets of [OIII] emitters split by stellar mass. Our measurements (listed in Table $\ref{tab:MZR}$) extend the trend of an increasing [OIII]/H$\beta$ with decreasing mass at high-redshift \citep{Sanders20}, although this trend flattens at masses below $10^9$ M$_{\odot}$ and suggestively turns over below $\lesssim10^8$ M$_{\odot}$. Our detection-rate of galaxies with low [OIII]/H$\beta\lesssim3$ ratios is highest at intermediate masses, which highlights significant scatter in these line-ratios at fixed mass. We discuss these objects in \S $\ref{sec:discus_metal}$.

The interpretation of the trend between [OIII]/H$\beta$ and mass is not straightforward as the [OIII]/H$\beta$ ratio is sensitive to variations in both the excitation state and the gas-phase metallicity. The relation between the gas-phase metallicity and [OIII]/H$\beta$ is double-valued with a peak at 12+log(O/H)$\approx7.7$ and [OIII]/H$\beta\approx6$ \citep[e.g.][]{Bian2018,Curti22,Nakajima22}. To illustrate this effect, the purple line in Fig. $\ref{fig:MEX}$ shows the expected relation between [OIII]/H$\beta$ and mass assuming the \cite{Bian2018} strong-line calibration derived in local analogues of high-redshift galaxies and the mass-metallicity relation (MZR) at $z\approx6$ in the FIRE simulation \citep{Ma16}\footnote{Similar to e.g. \cite{Curti22}, we rescaled the FIRE MZR to match the absolute normalisation of the measured MZR at $z\approx3$ by \cite{Sanders21}, but apply the redshift evolution measured in the simulation \citep{Ma16}.}, which shows a similar behaviour as observed in our stacks. This comparison implies that the metallicities of our [OIII] emitters with mass $\approx10^{8-9}$ M$_{\odot}$ are close to the metallicity where the relation between [OIII]/H$\beta$ and metallicity peaks. We thus conclude that the extremely high typical [OIII] EWs in the majority of our sample are found in galaxies with ISM conditions that give rise to the maximum [OIII] luminosity at given H$\beta$ luminosity (and thus, plausibly, at a given star formation rate).  

\begin{figure}
    \centering
    \includegraphics[width=8.3cm]{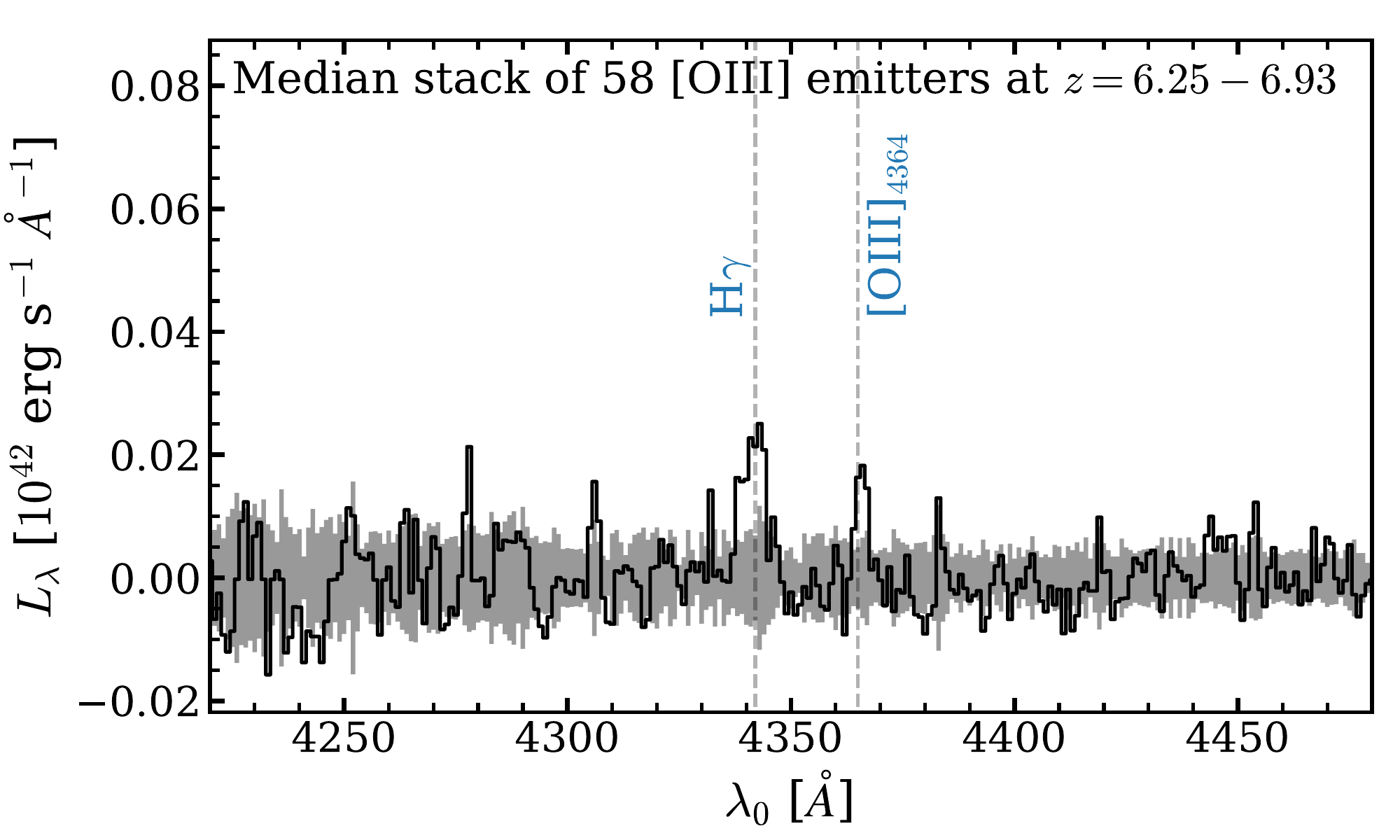}
    \caption{Median stacked 1D spectrum of the sub-set of [OIII] emitters at $z=6.25-6.93$ for which we have complete coverage of H$\gamma$ and all redder lines, zoomed in on H$\gamma$ and [OIII]$_{4364}$. The faint detections are at S/N of 6.5 and and 5.3, respectively, but they allow us to constrain the typical dust attenuation and electron temperature.  }
    \label{fig:O3_4364}
\end{figure}

Surprisingly, we detect the temperature sensitive [OIII]$_{4364}$ line in the stacked spectrum of [OIII] emitters at $z>6.25$ (where we have full coverage for this line and H$\gamma$), see Fig. $\ref{fig:O3_4364}$. After correcting the [OIII]$_{5008}$/[OIII]$_{4364}$ ratio for dust attenuation using the measured Balmer decrement, we measure an electron temperature $T(O^{++})=2.2\pm0.2\times10^4$ K using \texttt{PyNeb} \cite{Luridiana15} assuming an electron density of 300 cm$^{-3}$ \citep[e.g.][]{Sanders16,Curti22}. While this is a fairly high electron temperature, it is not extreme compared to other recent measurements in high-redshift galaxies or extremely low metallicity galaxies in the nearby Universe \citep[e.g.][]{Katz22} and in line with early galaxies showing highly ionised and heated interstellar medium. 

The measurement of the electron temperature allows a direct estimate of the $O^{++}$ abundance. Following the methodology outlined in \cite{Pilyugin06} and propagating the uncertainties in the various line-ratios, we measure 12+log$_{10}$($O^{++}$)$_{\rm direct}=7.21\pm0.08$. We need to assume an [OIII]/[OII] ratio to estimate the total gas phase metallicity as our spectra do not cover the [OII]$_{3727,3729}$ line.  Based on empirical scalings of the [OIII]/[OII] ratio with line EW and electron temperature \citep[e.g.][]{Reddy18b,Katz22} we assume [OIII]/[OII]$=8\pm3$. We then find a slightly higher total gas phase metallicity of 12+log$_{10}$(O/H)$_{\rm direct}=7.38\pm0.09$. This estimate is within 1$\sigma$ agreement with the estimate based on [OIII]/H$\beta$ using the strong-line calibration from \cite{Bian2018} assuming it is on the lower metallicity branch, which results in 12+log$_{10}$(O/H)$_{\rm O3H\beta}=7.25\pm0.08$.

\begin{table}[]
    \centering
    \caption{The combined rest-frame EW(H$\beta$+[OIII]$_{4960,5008}$) and [OIII]$_{5008}$/H$\beta$ line-ratios in stacks of four subsets of our sample (with H$\beta$ coverage at $z>5.5$) split by stellar mass. Uncertainties for the EW are the 16-84 percentiles of the distribution of EWs within each subset, while uncertainties on line-ratios are measurement errors. We convert these line-ratios to a gas-phase metallicity based on the strong-line calibration by \cite{Bian2018}, see text. Logarithms are to base 10. } 
    \begin{tabular}{cccc}
     log(M$_{\star}$/M$_{\odot}$)   & EW$_{0, \rm H\beta+[OIII]}$/{\AA} & [OIII]/H$\beta$ & 12+log(O/H) \\ \hline
 7.5 & $1870^{+1200}_{-590}$ & $5.3^{+0.9}_{-0.8}$ & $7.29^{+0.15}_{-0.14}$ \\
  8.2 & $980^{+670}_{-240}$ & $6.7^{+0.6}_{-0.6}$ & $7.77^{+0.22}_{-0.20}$ \\
 8.9 &$690^{+340}_{-300}$ & $6.3^{+0.9}_{-0.7}$ & $8.05^{+0.12}_{-0.35}$ \\
 9.5 & $410^{+200}_{-100}$ & $5.4^{+0.5}_{-0.4}$  & $8.19^{+0.06}_{-0.06}$ \\
    \end{tabular}
    \label{tab:MZR}
\end{table}

\begin{figure*}
    \centering
    \includegraphics[width=13.3cm]{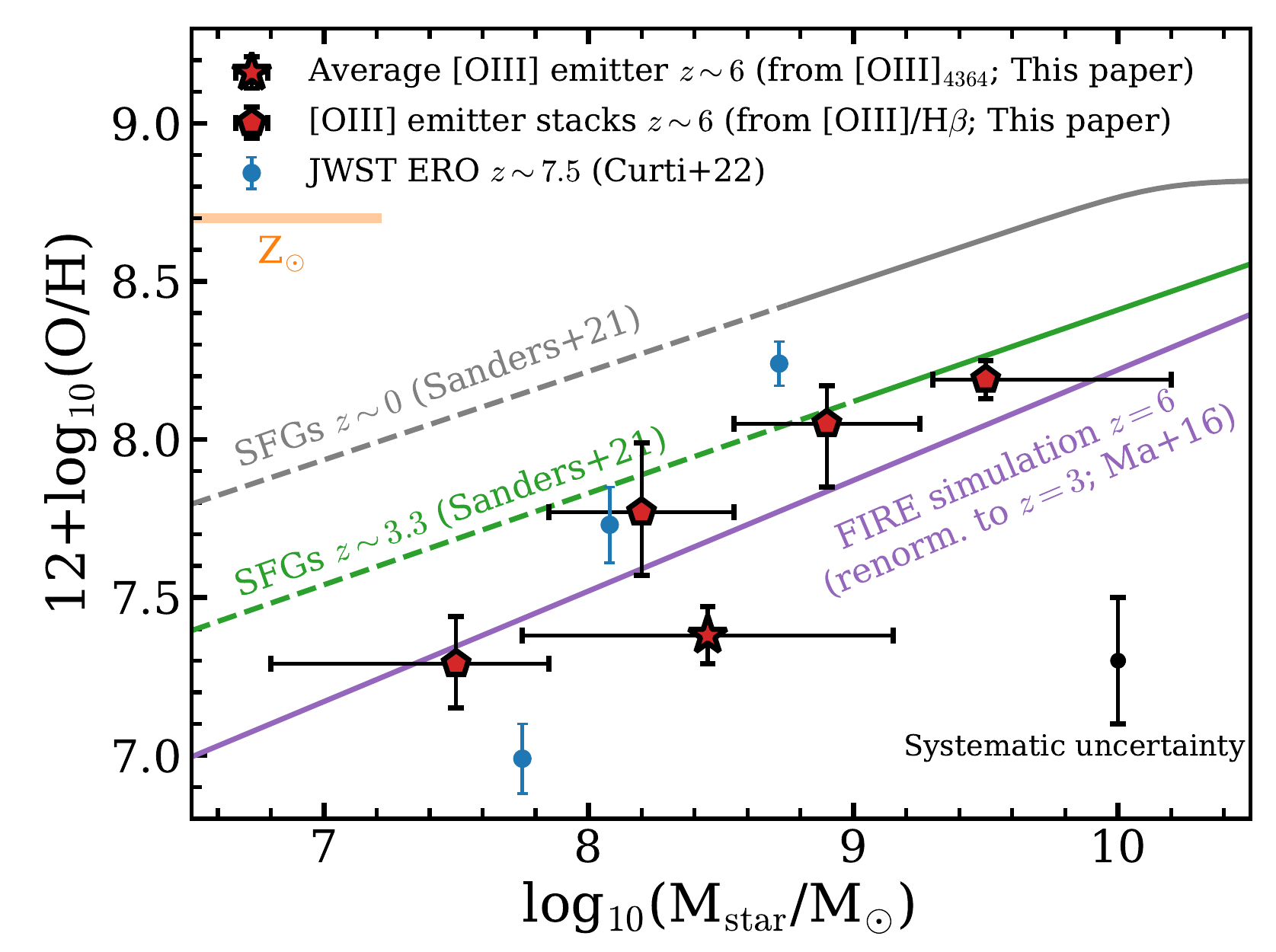}
    \caption{The relation between gas-phase metallicity and stellar mass at $z\sim6$ for galaxies and stacks with direct $T_e$-based or strong-line based metallicity estimates. The horizontal errors show the boundaries of the subsets (pentagons) and the 16-84th percentiles of masses in the full stack (star), respectively. The solar metallicity \citep{Asplund2009} is highlighted in orange. We highlight the systematic uncertainty between different strong-line calibrations in the bottom-right. We find that recent measurements in individual galaxies \citep{Curti22} roughly scatter around the metallicity of our median stack of $z\sim6.5$ [OIII] emitters. The average metallicity is slightly higher than expectations from the FIRE simulation \citep{Ma16}, once these have been rescaled to match the mass-metallicity relation at $z\approx3$ \citep{Sanders21}.}
    \label{fig:MZR}
\end{figure*}

Now, under the assumption that the \cite{Bian2018} calibration is also applicable for our stacks in subsets of mass, we estimate gas-phase metallicities for these subsets and list them in Table $\ref{tab:MZR}$. Based on the relation between [OIII]/H$\beta$ and mass that is shown in Fig. $\ref{fig:MEX}$, we assume that our two most massive subsets are on the higher metallicity branch of the [OIII]/H$\beta$ - metallicity relation, while the other subsets are on the lower branch. In Fig. $\ref{fig:MZR}$ we compare the $T_e$-based metallicity measurement of our median stack and our strong-line method based measurements in subsets of mass to recent measurements in early {\it JWST} spectra \citep[e.g.][]{Curti22} and (rescaled) expectations from hydrodynamical simulations \citep{Ma16}. Our strong-line method estimates are suggestive of a strong correlation between metallicity and mass, in line with expectations from the simulations. Our results extend the redshift evolution of a decreasing metallicity at fixed mass established from $z\approx0-3$ \citep[e.g.][]{Sanders21}, and our measurement for the sample averaged metallicity (from the $T_e$-method) is in line with the redshift evolution expected from these simulations. 

We however note that the applicability of the specific strong-line calibration is uncertain. For example, our direct-method based metallicity for the full stack of galaxies at $z>6.25$ is significantly lower than the strong-line based metallicity of subsets with similar mass as the median mass of the full sample. While this could be due to calibration uncertainties, we remark that line-ratios measured in median stacks of samples with a relatively wide variation in line-ratios -- especially with values around the peak of a double-valued relation -- may be challenging to interpret. Further, if we would use the \cite{Nakajima22} calibration for sources with high EWs, we find that the metallicities of our two lowest mass stacks would increase by $\approx0.2$ dex yielding milder redshift evolution at $z\gtrsim3$ and a shallower slope. Future follow-up targeting fainter lines such as [OIII]$_{4363}$ and ongoing efforts to re-calibrate strong line methods at high-redshift will be able to relieve some of these caveats. Further, larger samples from the full EIGER data will enable stacks in subsets with smaller dynamic range.

\section{Discussion and Outlook} \label{sec:discuss}
Here we will discuss our results and provide an outlook of some results that can be anticipated with the observations of the full EIGER sample in six quasar fields.

\subsection{Strong [OIII] lines in typical $z\sim6$ galaxies}  \label{sec:ubiq}
Our large spectroscopic sample of [OIII] emitting galaxies at $z=5.33-6.93$ confirms that strong rest-frame optical emission-lines are abundant in distant galaxies, with typical EWs of $\approx850$ {\AA} at UV luminosities M$_{\rm UV}\approx-19.5$ (\S $\ref{sec:EW}$). However, given that our sample is a line-selected sample, this result may not be surprising.

Our measured EWs are in broad agreement with inferences for UV-selected galaxies from SED modeling of the rest-optical light captured in {\it Spitzer}/IRAC or {\it JWST} filters \citep[e.g.][]{Smit14,Roberts-Borsani16a,Endsley21,Endsley22b}, see Fig. $\ref{fig:EWmass}$. Furthermore, as shown in Fig. $\ref{fig:MUVLO3}$, the relation between the [OIII] line luminosity and the UV luminosity in our sample is only slightly lower compared to the inferred relation in UV-selected galaxies at $z\sim8$ \citep{debarros19}. These comparisons imply that these strong lines are {\it typical} in UV-selected galaxies and that there are no significant differences between samples that are selected either through the Lyman break or [OIII] line-emission. 

\begin{figure}
    \centering
    \includegraphics[width=8.3cm]{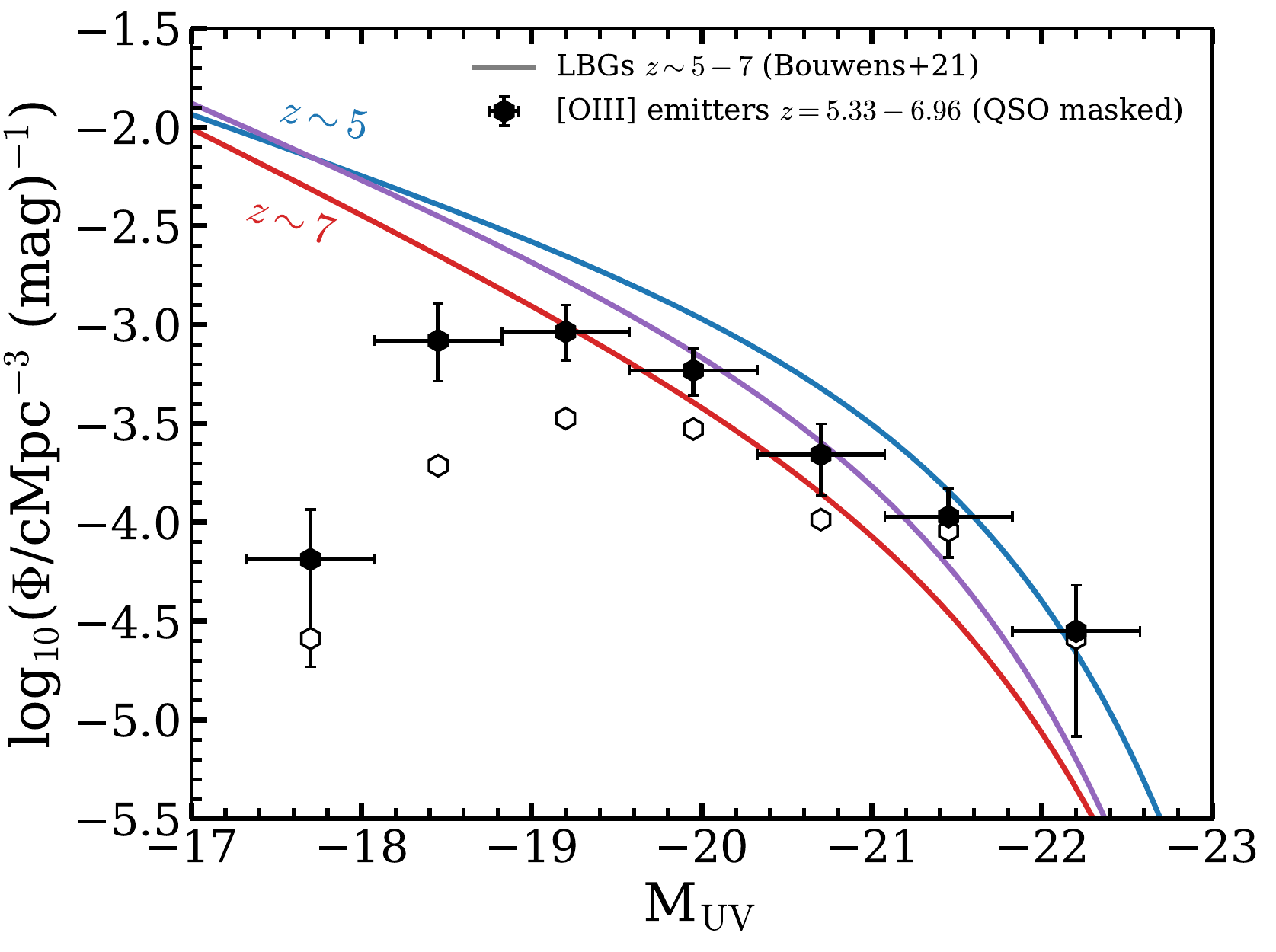}
    \caption{The field UV luminosity function of our sample of [OIII] emitters at $z=5.33-6.96$, masking the redshift around the quasar (hexagons). Open hexagons show the number densities without correcting for completeness and the maximum volume of each individual source. The number densities match the UV luminosity function of UV-selected galaxies at $z\sim6$ by \cite{Bouwens21} at $-21.0<$M$_{\rm UV}<-19.5$. This demonstrates that the strong lines are typical for most $z\sim6$ galaxies. At fainter UV luminosities, our line-selected sample only picks up the upper end of the EW distribution. Cosmic variance is likely important at the bright end. }
    \label{fig:UVLF}
\end{figure}

We derive the UV luminosity function of our sample of [OIII] emitters following the methods outlined in \S $\ref{sec:LF}$. As shown in Fig. $\ref{fig:UVLF}$, we find that the number densities of [OIII] emitters agree relatively well with the UV LF of LBGs measured by \cite{Bouwens21} at $z\sim6$, at least at luminosities in the range M$_{\rm UV}=-19.5$ to M$_{\rm UV}=-21.0$. Similar to our [OIII] LF, we have masked the redshift around the targeted quasar in order not to be biased by its over-density. The resemblance of these number densities, despite that our target selection is completely different from the standard LBG selections \citep[e.g.][]{Bouwens21}, further demonstrate that the strong [OIII] emission-lines are typical in LBGs. At fainter UV luminosities, the number densities of [OIII] emitters is significantly lower than the UV LF. This can be understood in the context of the [OIII] EW distribution: for these faint galaxies, our emission-line survey only picks up galaxies with the most extreme EWs, which are no longer representative for the full sample at those UV luminosities. At the bright end, we note that cosmic variance may be particularly important. Three of the four galaxies with M$_{\rm UV}\lesssim-22$ are part of a large over-density at $z\approx6.77$ \citepalias{SurveyPaper}. The addition of the five extra quasar fields from the EIGER program will be particularly useful to constrain the shape of the bright-end of the LF.

\begin{figure*}
    \centering
    \includegraphics[width=11.3cm]{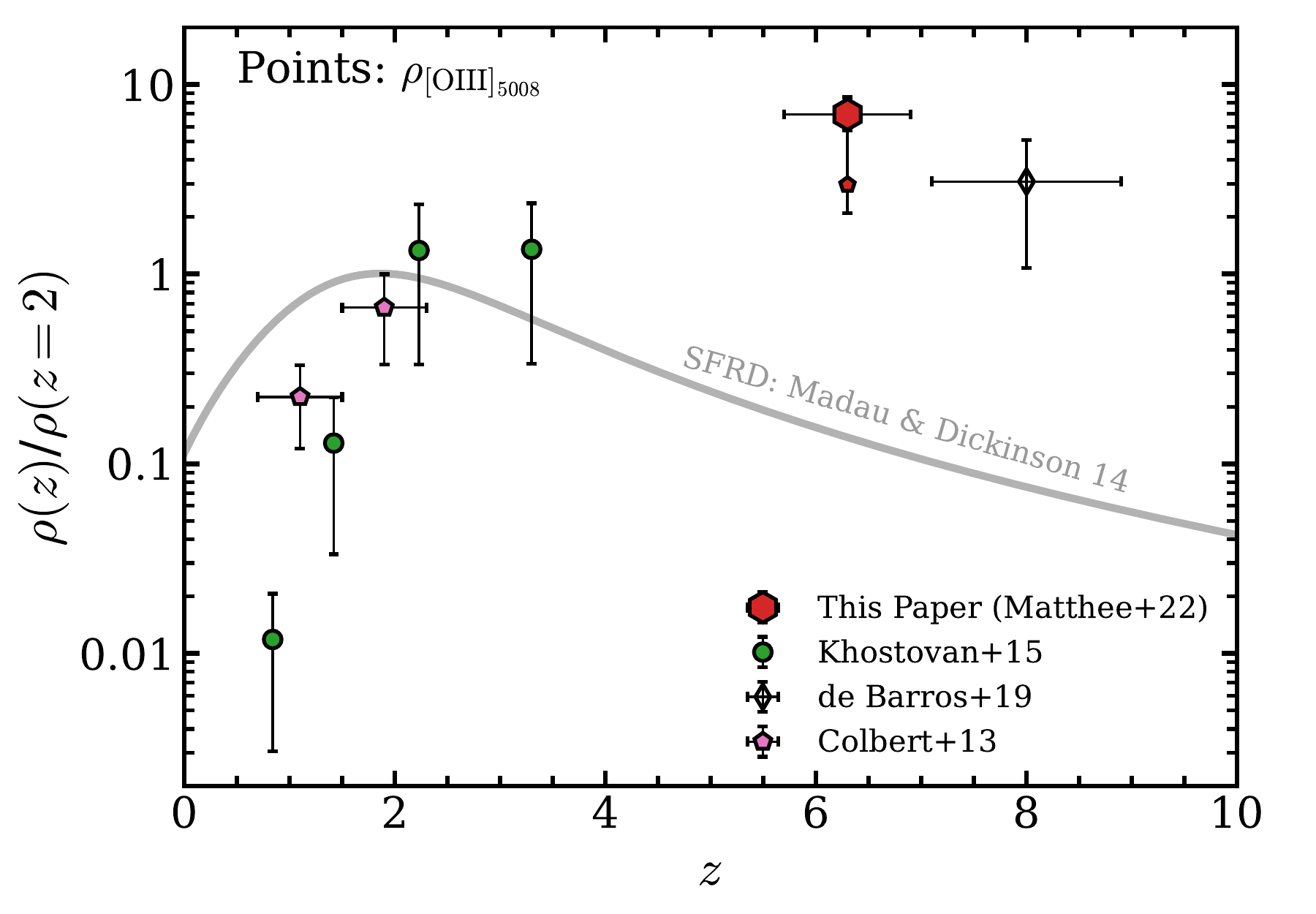}
    \caption{The field [OIII]$_{5008}$ luminosity density (integrated down to a limiting luminosity of $10^{42.2}$ erg s$^{-1}$) as a function of redshift, scaled to the density at $z=2$. Our results based on {\it JWST} spectroscopy are shown as a red hexagon (assuming a faint-end slope $\alpha=-2$, see \S $\ref{sec:LF}$). The small red pentagon shows our results for our fit with varying faint-end slope (best-fit $\alpha=-0.2$), showing that this does not significantly affect out results. The SED infered [OIII] luminosity density at $z\sim8$ is shown as open diamond \citep{debarros19}. Green circles show the narrow-band luminosity densities measured by \citep{Khostovan15} and the pink data-points show results from the {\it HST} grisms at lower redshifts \citep{Colbert13}. The grey line shows the evolution of the cosmic star formation rate density from \cite{madau&dickinson}, also scaled to $z=2$.}
    \label{fig:O3dens}
\end{figure*}

Based on our spectroscopically measured [OIII] luminosity function and earlier measurements based on photometry or {\it HST} grism spectroscopy at lower redshifts, we now compare the evolution of the [OIII] luminosity density $\rho_{\rm [OIII]}$ to the evolution of the cosmic star formation rate density from \cite{madau&dickinson} in Fig. $\ref{fig:O3dens}$. The [OIII] luminosity density evolves rather differently from the star formation rate density, with a significantly earlier peak in cosmic time. Whether the $\rho_{\rm [OIII]}$ indeed peaks at $z\approx6$ as our results suggest, or perhaps between $z\approx4-6$, needs to be verified with future {\it JWST} observations at lower redshifts. 

Despite an order of magnitude decline in the cosmic SFR density from $z=2-6$, we find that $\rho_{\rm [OIII]}$ in fact increases from $z=2$ to $z=6$ by a factor $2 - 5$. Our explanation for this trend is a conspiracy between the evolving star formation rate density, the mass function of star-forming galaxies and the chemical enrichment, dust attenuation and photo-ionization conditions. As the stellar metallicity decreases towards high-redshift \citep[e.g.][]{Cullen19,Kashino22}, the ionizing efficiency of stellar populations increases (see also Fig. $\ref{fig:Xion}$) yielding higher line luminosities at fixed SFR in general. Further, the harder spectra of these stellar populations lead to higher excitation conditions \citep[e.g.][]{Steidel16} that primarily result in an increasing [OIII] luminosity as long as the gas-phase metallicity is above 12+log(O/H)$\gtrsim7.5$ (Fig. $\ref{fig:MEX}$). Finally, most star formation in galaxies in the $z\approx2$ Universe occurs in objects with a relatively high mass $\sim10^{10}$ M$_{\odot}$ \citep[e.g.][]{Davidzon17,Chruslinska19,Behroozi19} which have significantly higher metallicity \citep{Sanders21}, dust attenuation \citep{GarnBest10} and lower ionizing photon production efficiency \citep[e.g.][]{Atek22a}. Therefore, while the rate at which galaxies form stars is lower at higher redshifts, the lower metallicities of the stellar atmospheres and interstellar media make these galaxies in comparison much more luminous in [OIII] (see also Fig. $\ref{fig:EWmass}$). In this picture, it is indeed not a surprise that we measure a typical [OIII]/H$\beta$ that is close to the maximum expected in star-forming galaxies ([OIII]$_{5008}$/H$\beta \approx7$; e.g. \citealt{Bian2018}).

At some point, these favourable conditions for [OIII] emission should no longer be balanced by the declining star formation rate density. Further, the [OIII] luminosity decreases at even lower metallicities than the ones that characterise our sample (as indicated by our stacking results in Fig. $\ref{fig:MZR}$. These processes combined should lead to a rapid decline in $\rho_{\rm [OIII]}$ at $z>6$ that could potentially probed by future {\it JWST} observations in redder NIRCam filters. Likewise, it is plausible that the [OIII] luminosity declines rapidly in a significant fraction of galaxies with masses below $<10^7$ M$_{\odot}$, which would lead to a flattening of the [OIII] LF towards fainter luminosities, and a less dramatic difference between the evolution of $\rho_{\rm [OIII]}$ and the star formation rate density once we would integrate to fainter luminosity limits. Characterising such flattening of the [OIII] LF should be performed with deeper complete emission-line surveys.

\subsection{The role of galaxies during reionization}
A prime goal of our EIGER survey is to observe the role of galaxies in the reionization of the Universe, in particular using the cross-correlation between galaxies and Ly$\alpha$ forest data in the quasar spectra. The ability to simultaneously measure the ionizing emissivity $\xi_{\rm ion}$ for these galaxies is promising. Generally, our measured log$_{10}(\xi_{\rm ion}$/Hz erg$^{-1})=25.31^{+0.29}_{-0.16}$ is somewhat higher than the canonical values that have been assumed when modeling the role of galaxies in the reionization of the Universe \citep[e.g.][]{Robertson13}. High values of $\xi_{\rm ion}$ imply that more modest values of the escape fraction of ionising photons are needed for galaxies to reionize the Universe by $z\approx6$ \citep[e.g.][]{Davies21}, or that the contribution from very faint galaxies is minor \citep[M$_{\rm UV}>-17$ e.g.][]{MattheeNaidu22}.

As shown in \citetalias{SurveyPaper}, we find clear indications of an excess Ly$\alpha$ and Ly$\beta$ transmission at a distance of $\approx5$ cMpc around [OIII] emitters at $z\approx6$ in this single quasar sight-line. This demonstrates the local enhancement in the ionising emissivity around galaxies and supports an important role of galaxies in the end stages of cosmic reionization. With the full EIGER data, our simultaneous measurement of $\xi_{\rm ion}$ may enable us to constrain the escape fraction and contribution of faint unseen galaxies using a larger sample of quasar sight-lines \citep[e.g.][]{Kakiichi18,Meyer20}.

\subsection{The most metal poor galaxies in our sample} \label{sec:discus_metal}
\begin{figure}
    \centering
    \includegraphics[width=8.5cm]{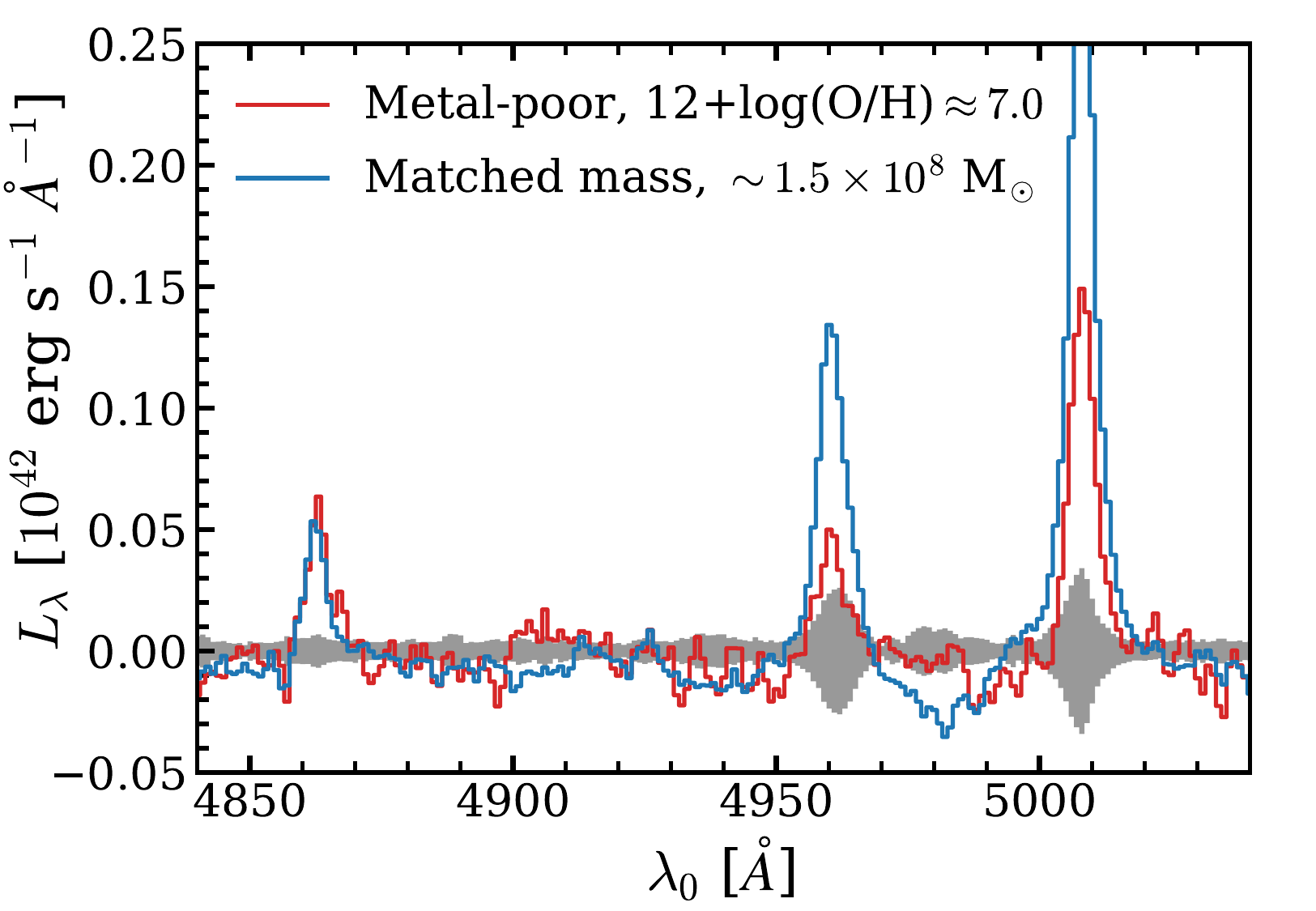}
    \caption{The stacked spectrum of the five metal-poor low mass galaxies with the lowest [OIII]/H$\beta$ ratios (where H$\beta$ was detected with S/N$>5$; red) galaxies compared to the stacked spectrum of galaxies with comparable mass (blue).}
    \label{fig:metalpoor}
\end{figure}

The early results on the gas-phase metallicities in $z\sim6-7$ galaxies (\S $\ref{sec:metal}$) suggest that there are significant metallicity variations. As can be seen in Fig. $\ref{fig:MEX}$, the five objects with the lowest [OIII]/H$\beta$ ratios ($\lesssim3$) appear among the intermediate masses probed by our sample. We show the stacked spectrum of these five galaxies in Fig. $\ref{fig:metalpoor}$, where we compare it to the stacked spectrum of all galaxies with similar mass. We need to rely on strong-line calibrations to estimate the gas-phase metallicities of these galaxies, which are particularly uncertain at these [OIII]/H$\beta$ ratios if one assumes that they are on the lower metallicity branch (see e.g. \citealt{Curti22}). Using the two calibrations that encapsulate the range of possible values \citep{Bian2018,Nakajima22}, we find a metallicity in the range 12+log(O/H)=$(6.8-7.1)\pm0.1$ (i.e. 1-2 \% solar). These metallicities are $\approx0.6$ dex lower than the metallicity estimated for the comparison sample (Table $\ref{tab:MZR}$), confirming the significant scatter in metallicities at fixed mass. A more detailed investigation of the properties of these systems requires further spectroscopic follow-up observations that can verify the metallicities and characterise the full SED. The fact that these objects can be identified in NIRCam WFSS data is promising for statistical studies of the properties of low metallicity galaxies.

\subsection{On the efficiency of NIRCam WFSS surveys}
Finally, with the aim of the community planning future {\it JWST} observations, we reflect on the efficiency and challenges of NIRCam WFSS surveys based on our first analyses. 

The main limitation in WFSS is the spectral contamination and the association between spectral features and galaxies. The latter is particularly challenging in the case of a single line detection with high equivalent width where the continuum trace is not visible. Typically, both issues are mitigated by observing the field with multiple (ideally orthogonal) dispersion directions \citep[e.g.][]{Brammer12}. However, obtaining multiple orthogonal dispersion directions with NIRCam is relatively inefficient. Nevertheless, we find that several factors significantly mitigate these issues, specifically when searching for emission-line doublets. These are the high spectral resolution ($R\sim1500$) and the relatively flat continuum spectrum of the main foreground population at $3.5~\mu$m such that continuum contamination can be cleaned out very efficiently thanks to the high contrast of narrow emission-lines (see Fig. $\ref{fig:data_illustration}$). The use of doublets (or sets of relatively nearby lines as H$\beta$+[OIII]) further allows a precise estimate of the observed wavelength of the lines. This facilitates the galaxy - line association significantly, even in the 80 \% of our survey area that is only observed with a single dispersion direction. For galaxies with single lines (such as H$\alpha$ emitters in practice are), the correct line-galaxy association in single-dispersion WFSS data either requires morphological information and/or prior constraints on the redshift from broadband colors.

Generally, the main strengths of a NIRCam/WFSS survey are spectra for a complete coverage of all objects within the field of view without photometric pre-selection, the ability to perform spatially resolved studies and robustness against uncertainties in slit-losses or uncontrolled contamination \citep[e.g.][]{Maseda19}. Indeed, our survey has efficiently yielded redshifts for $>100$ [OIII] emitters at $z=5.3-7$ already in 18.5 hours of {\it JWST} observing time including overheads. A similar number of H$\alpha$ emitters at $z=3-5$ (to be explored in future analyses) and several spatially resolved lower redshift $z\approx1-3$ lines from the Paschen and Brackett series, and infrared lines as HeI and FeII  \citep[e.g.][]{Brinchmann22} are further lurking in the data. Additionally, thanks to the design of NIRCam, {\it JWST} can simultaneously obtain very sensitive $<2.4\mu$m imaging in the same field of view. All things combined, we conclude that thanks to the brightness of the [OIII] doublet in high-redshift galaxies, NIRCam WFSS surveys will be particularly efficient in mapping the complete distribution of star-forming galaxies with M$_{\rm star}\gtrsim10^7$ M$_{\odot}$ from $z\approx3-9$ in the field of view.

\section{Summary} \label{sec:summary}
It is a key aim to characterise the physical conditions of high-redshift galaxies in order to understand early galaxy formation and probe the epoch of reionization. Here, we use the first deep $3.5\mu$m wide field slitless spectroscopic (WFSS) {\it JWST} observations of the EIGER program to analyse the properties of 117 spectroscopically confirmed [OIII] emitters at $z=5.33-6.93$ in the field of the luminous quasar J0100+2802. The F356W WFSS observations are complemented with NIRCam imaging in the F115W, F200W and F356W filters. These first deep observations demonstrate the excellent performance of WFSS using {\it JWST}/NIRCam in identifying large complete samples of distant galaxies \citepalias[see][]{SurveyPaper}. Our main results are summarised in the following points:

\begin{itemize} 
\item Using an automated search algorithm that detects [OIII]$_{4960,5008}$ and [OIII]+H$\beta$ pairs with S/N $\geq3$ in the grism data we identify a total of 133 resolved [OIII] emitting components, of which 68 show additional H$\beta$ detections and two show H$\gamma$.

\item We find that a large number of [OIII] emitters are closely separated, with a strong excess clustering at $<2''$. All such close pairs are physically associated ($|\Delta z| \lesssim 600$ km s$^{-1}$). In order not to have our sample selection depend on the spatial resolution, deblending parameters and our orientation to the objects, we merge all closely separated pairs in so-called systems and threat them as one of the 117 galaxies that constitute our sample.
\item By self-consistently modeling the nebular and stellar components that contribute to the observed spectral energy distributions and fitting this to the spectral and photometric data, we find that the galaxies in our sample are characterised by relatively young ages ($\approx100$ Myr) and low dust attenuation (E$(B-V)\approx0.1$). Our sample spans a wide UV luminosity range M$_{\rm UV}=-17.7 $ to M$_{\rm UV}=-22.3$ and the masses are typically small ($\approx2\times10^8$ M$_{\odot}$), while in total spanning three orders of magnitude ($10^{6.8-10.1}$ M$_{\odot}$). 

\item We spectroscopically measure an average H$\beta$+[OIII] EW of $948^{+192}_{-138}$ {\AA} for UV bright galaxies (M$_{\rm UV}<-20.5$). This confirms that such extreme rest-frame optical EWs that are very rare in the local Universe ($<1$ \% of SDSS) are typical in the epoch of reionization. Our SED models suggest a typical increasing EW with decreasing stellar mass, increasing from an EW $\approx850$ {\AA} at the typical UV luminosity of M$_{\rm UV}\approx-19.6$ up to EWs $\approx3000$ {\AA} for our most extreme systems with masses M$_{\rm star}\sim10^7$ M$_{\odot}$.

\item We present the [OIII] luminosity function (LF) in the field (i.e. masking the quasar environment) at $z\sim6$, the first spectroscopic measurement at $z>1.5$. The luminosity function is slightly higher compared to measurements at $z\approx3$ and comparable to recent inferences at $z\sim8$ based on SED modeling. The UV LF of [OIII] emitters matches the UV LF of LBGs at $z\sim6$ relatively well, except for the bright and faint ends. This demonstrates that, in general, strong [OIII] emission is typical among LBGs. While the bright end is likely subject to significant cosmic variance, our lower number density for the UV faintest luminosities shows that we are only picking up the extreme tail of the EW distribution for those objects. 

\item The sensitive spectroscopy allows us to explore the physical conditions in our sample of $z\sim6$ galaxies using detections of [OIII]$_{4364}$ in stacks and H$\gamma$, H$\beta$ and [OIII]$_{4960,5008}$ in stacks and individual sources. We measure $\xi_{\rm ion}$, the ionizing photon production efficiency of galaxies to be $10^{25.31^{+0.29}_{-0.16}}$ Hz erg$^{-1}$, which is slightly higher than typically assumed in calculations of the reionization budget of star-forming galaxies. The galaxies are further characterised by little nebular dust attenuation E$(B-V)=0.14^{+0.16}_{-0.14}$ and a SFR$\approx5$M$_{\odot}$ yr$^{-1}$. 

\item Typically, our sample is characterised by a very high [OIII]/H$\beta$ ratio which suggests that the interstellar medium in the average galaxy in our sample has a high ionisation parameter and a gas-phase metallicity that optimises the [OIII]/H$\beta$ ratio. The [OIII]/H$\beta$ ratio varies non-monotonically with mass, peaking at $\sim10^8$ M$_{\odot}$ and declining slightly to both higher and lower masses. This behavior can be explained in the context of a mass-metallicity relation and the double-valued behaviour of [OIII]/H$\beta$ and metallicity. 

\item Our detection of [OIII]$_{4364}$ yields an average metallicity 12+log(O/H)=$7.38\pm0.09$ and supports the use of strong-line calibrations to derive the mass-metallicity relation at $z\sim6$, which slope and evolution compared to $z\sim3$ roughly matching expectations from hydrodynamical simulations. We further detect several intermediate mass galaxies with relatively low [OIII]/H$\beta$ suggesting that they are very metal poor, 1-2 \% solar, and demonstrating significant scatter at fixed mass.

\item We show that the strong [OIII] EWs in high-redshift galaxies lead to an [OIII] luminosity density that is a factor $\approx2-5$ higher at $z\sim6$ compared to the peak of the cosmic star formation rate density at $z\sim2$, despite the order of magnitude decline in cosmic star formation rate. As discussed in \S $\ref{sec:ubiq}$, we argue that this is due to a complex combination of a star formation rate that is increasingly dominated by lower mass galaxies towards higher redshifts. These galaxies have higher production efficiencies of ionising photons, lower gas-phase metallicity and lower dust attenuation compared to the more massive galaxies that dominate the star formation rate density at $z\approx2$. All these factors combined enhance the emerging [OIII] luminosity at $z\approx6$ compared to $z\approx2$.
\end{itemize}

The main implication of this paper for future observations is that the abundant strong [OIII] emission-lines from galaxies in the early Universe, combined with the ease with which continuum contamination can be removed, make NIRCam WFSS observations a highly efficient mode to spectroscopically map the galaxy distribution from the peak of star formation to the epoch of reionization. Main open questions are whether and when the faint-end slope of the [OIII] luminosity function flattens compared to the galaxy star formation rate function due to the increasingly lower metallicities at lower masses. The unanticipated sensitivity to detect fainter lines as H$\gamma$ and [OIII]$_{4364}$ are promising for the study of dust attenuation, star formation, gas-phase metallicity and the ionizing production efficiency of galaxies. The complete samples selected from the WFSS data will further allow clustering measurements and spatially resolved properties (colors, line-ratios) from the emission-line galaxies, which we will explore in future work. Simultaneous measurements of ionising production efficiency and the Ly$\alpha$ forest transmission in the full EIGER sample may offer promising perspective on the contributions of various galaxy populations to cosmic reionization.

\facilities{\textit{JWST}}

\software{
    \texttt{Python},
    \texttt{matplotlib} \citep{matplotlib},
    \texttt{numpy} \citep{numpy},
    \texttt{scipy} \citep{scipy},
    \texttt{Astropy}
    \citep{astropy1, astropy2}}
    
\begin{acknowledgments}
We thank an anonymous referee for a constructive report. We thank Norbert Pirzkal for advice on spectral extraction methods, the JAGUAR and \texttt{Mirage} teams \citep{Williams18,Hilbert19} for making extremely useful data-sets and sofware available to help developing our reduction and analysis pipelines and Rohan Naidu for useful comments on SED fitting.

This work is based on observations made with the NASA/ESA/CSA James Webb Space Telescope. The data were obtained from the Mikulski Archive for Space Telescopes at the Space Telescope Science Institute, which is operated by the Association of Universities for Research in Astronomy, Inc., under NASA contract NAS 5-03127 for \textit{JWST}. These observations are associated with program \# 1243.

All of the data presented in this paper were obtained from the Mikulski Archive for Space Telescopes (MAST) at the Space Telescope Science Institute. The specific observations analyzed can be accessed via \dataset[https://dx.doi.org/10.17909/yc3h-jn44]{http://dx.doi.org/10.17909/yc3h-jn44}.

This research made use of the open source Python package \texttt{Mirage}, the JWST Multi Instrument Ramp Generator \citep{Hilbert19}.
This research made use of Photutils, an Astropy package for
detection and photometry of astronomical sources \citep{Bradley22}.

This work has been supported by JSPS KAKENHI Grant Number JP21K13956 (DK). RS acknowledges support from NASA
award number HST-GO-15085.001.
\end{acknowledgments}

\bibliography{Biblio}
\bibliographystyle{apj}

\appendix 
\section{All H$\beta$ + [OIII] spectra} \label{appendix:1D}
\begin{figure*}
    \centering
    \includegraphics[width=16cm]{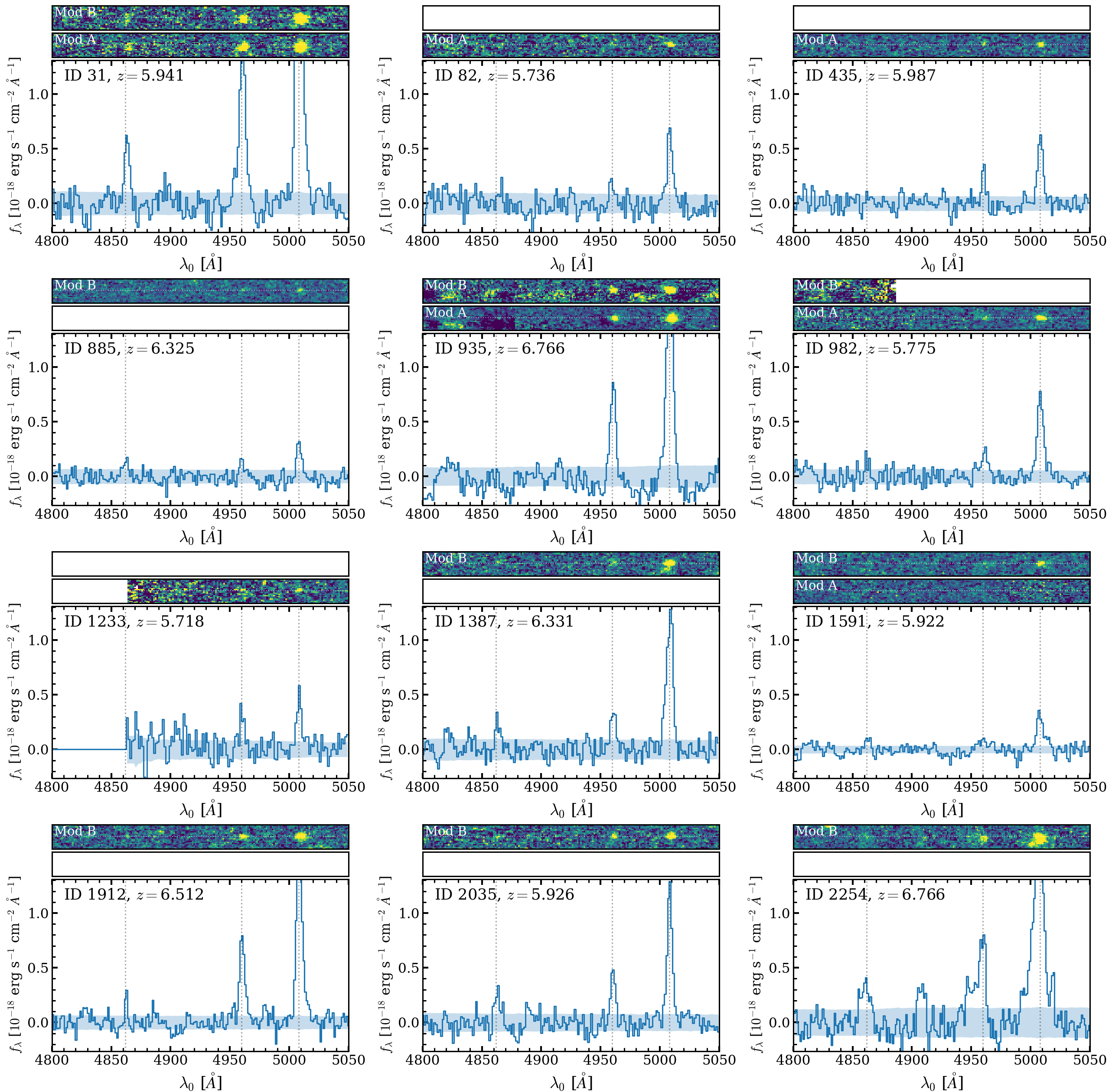}
    \caption{1D and 2D emission-line spectra of the [OIII] emitters in the sample centered on H$\beta$ and [OIII]$_{4960,5008}$. In each panel, we show the 2D grism spectrum in module A and module B (when covered). The 1D spectrum is the average of the optimally extracted 1D spectrum in each module. The blue shaded region shows the uncertainty.}
    \label{fig:spec1}
\end{figure*}

\begin{figure*}
    \centering
    \includegraphics[width=16cm]{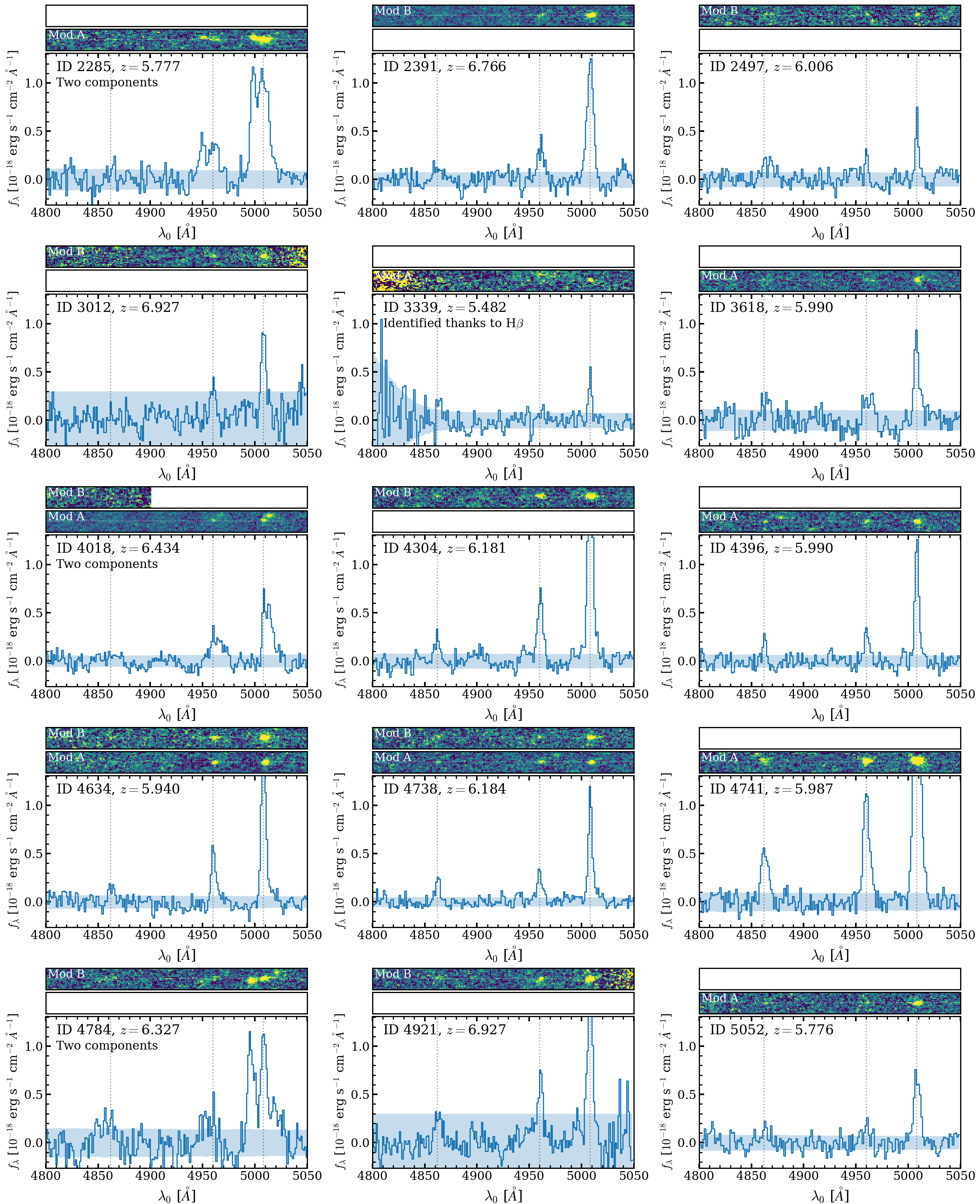}
    \caption{Fig. $\ref{fig:spec1}$, continued.}
    \label{fig:spec2}
\end{figure*}

\begin{figure*}
    \centering
    \includegraphics[width=16cm]{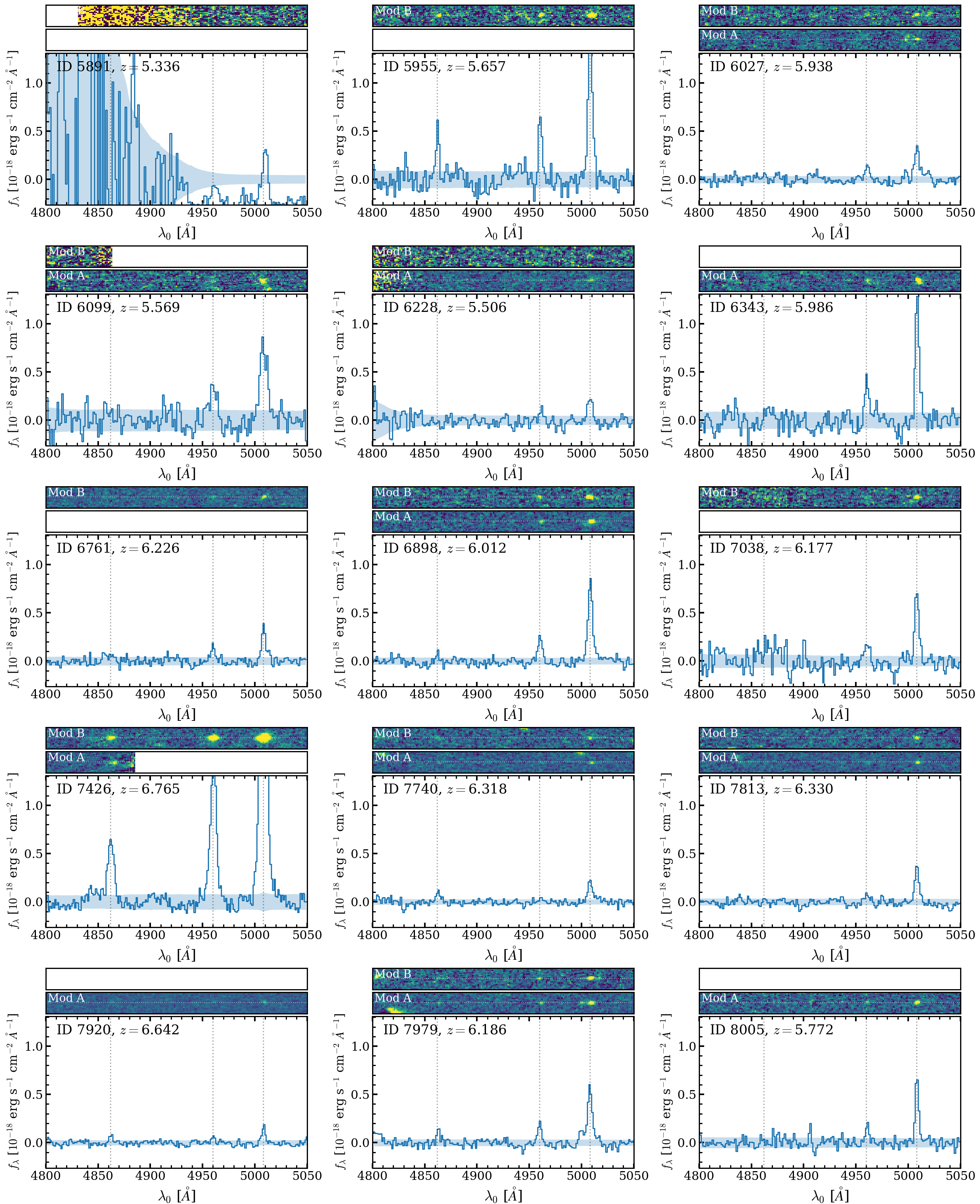}
    \caption{Fig. $\ref{fig:spec1}$, continued.}
    \label{fig:spec3}
\end{figure*}

\begin{figure*}
    \centering
    \includegraphics[width=16cm]{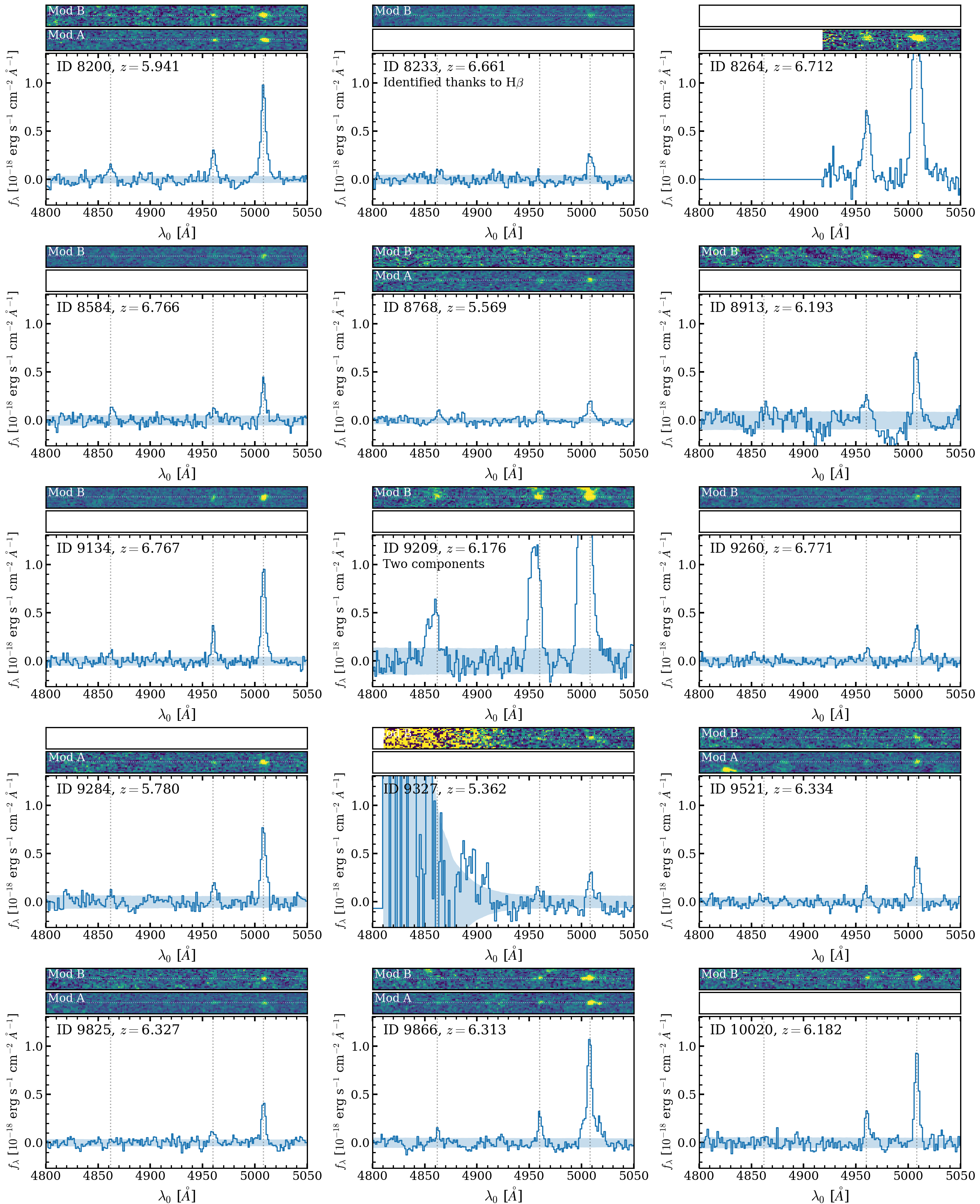}
    \caption{Fig. $\ref{fig:spec1}$, continued.}
    \label{fig:spec4}
\end{figure*}

\begin{figure*}
    \centering
    \includegraphics[width=16cm]{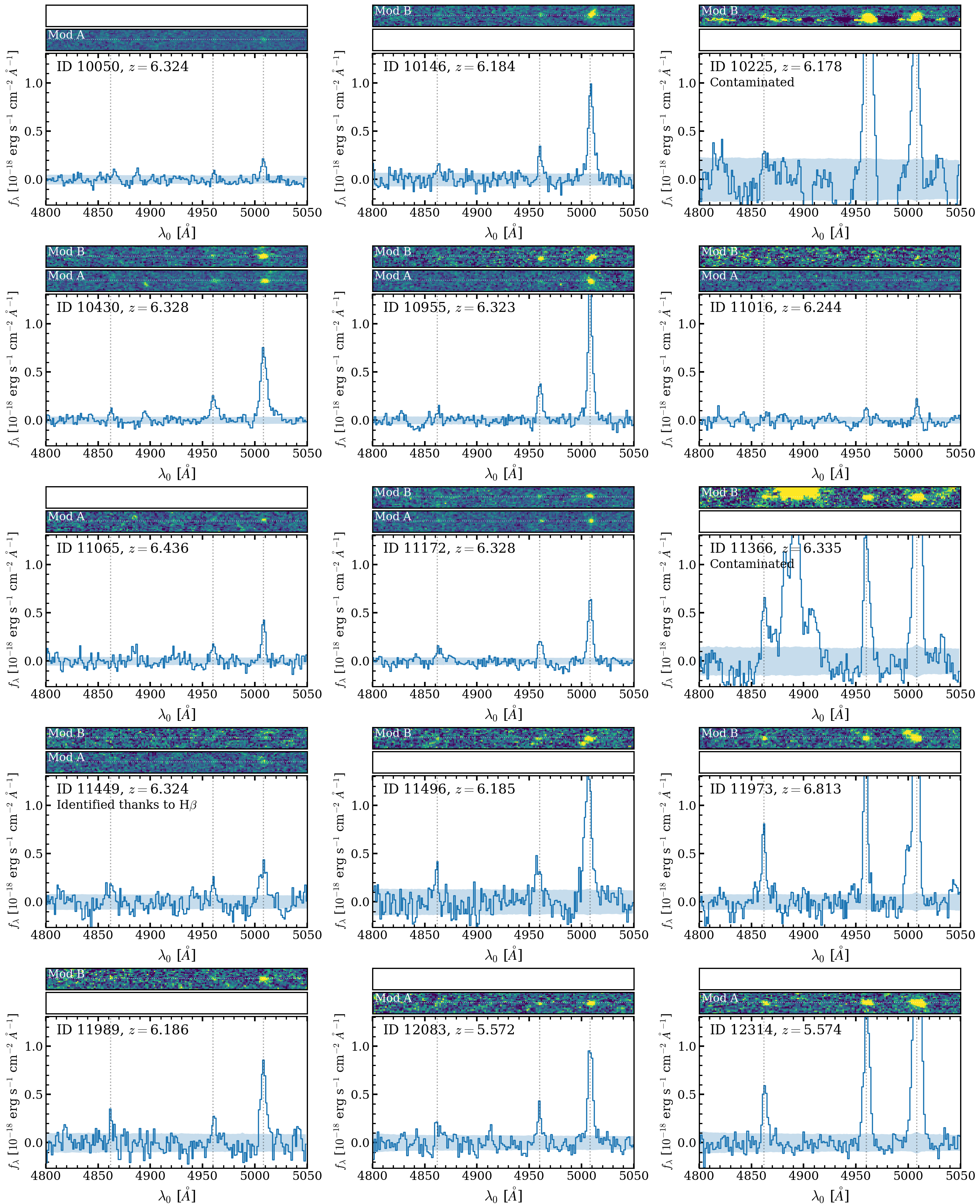}
    \caption{Fig. $\ref{fig:spec1}$, continued.}
    \label{fig:spec5}
\end{figure*}

\begin{figure*}
    \centering
    \includegraphics[width=16cm]{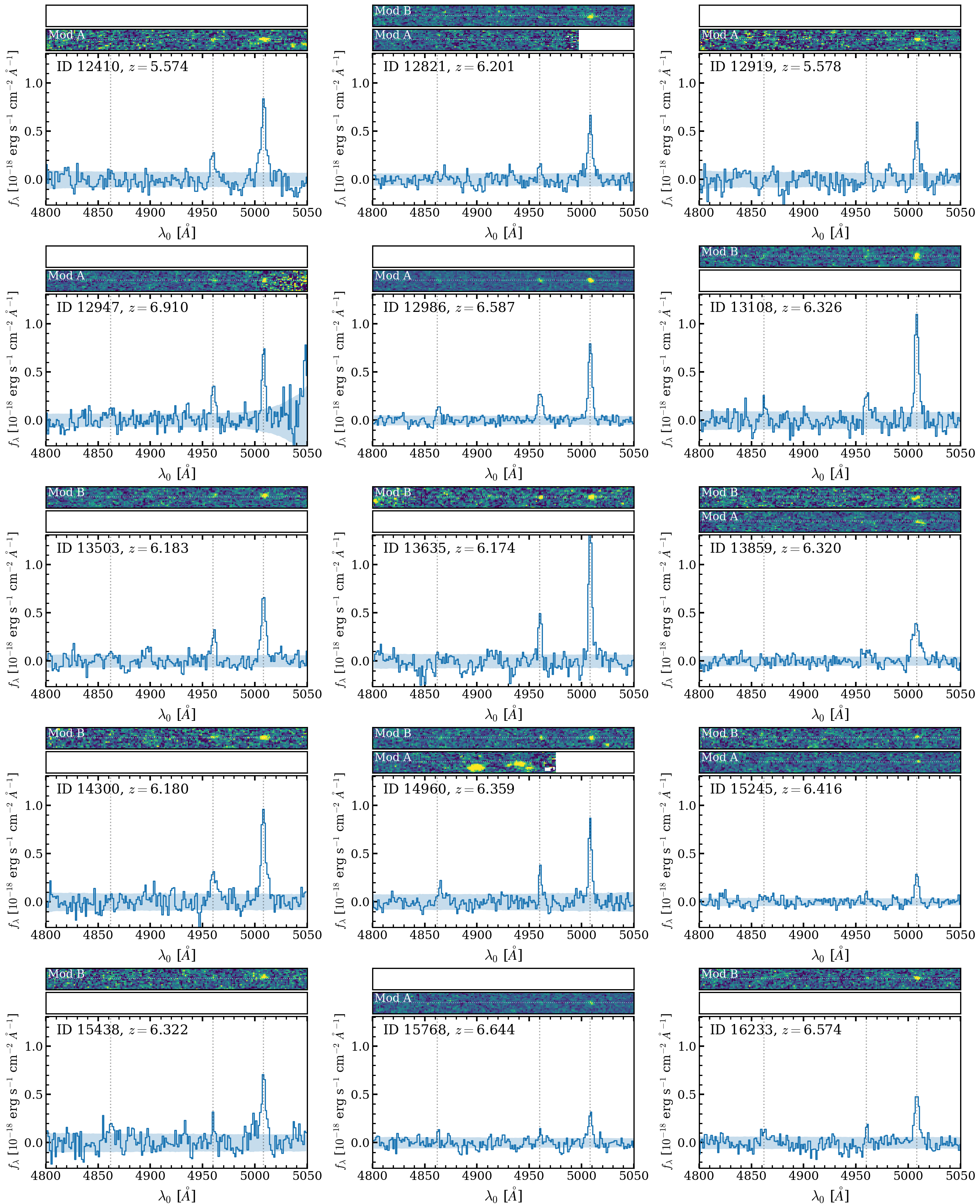}
    \caption{Fig. $\ref{fig:spec1}$, continued.}
    \label{fig:spec6}
\end{figure*}

\begin{figure*}
    \centering
    \includegraphics[width=16cm]{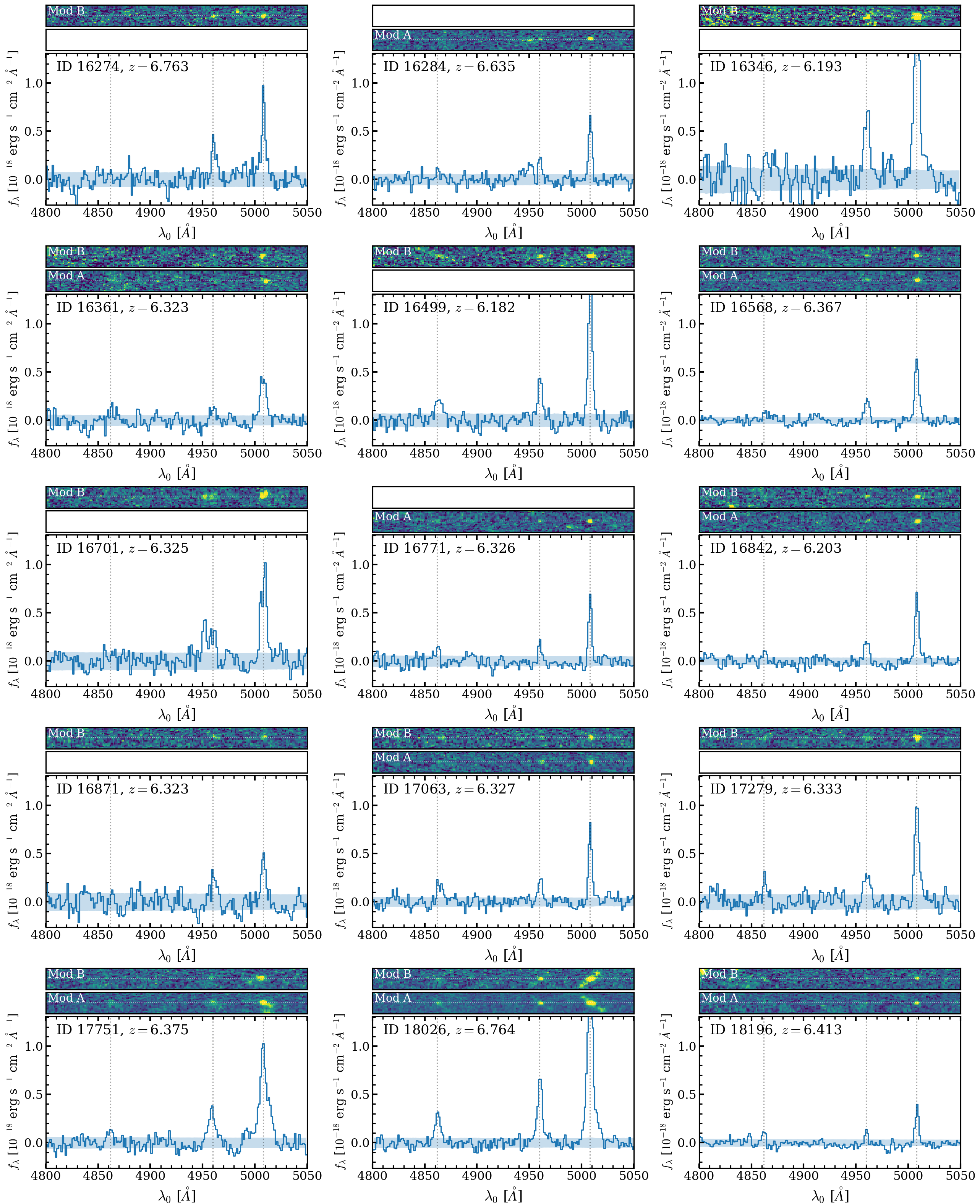}
    \caption{Fig. $\ref{fig:spec1}$, continued.}
    \label{fig:spec7}
\end{figure*}

\begin{figure*}
    \centering
    \includegraphics[width=16cm]{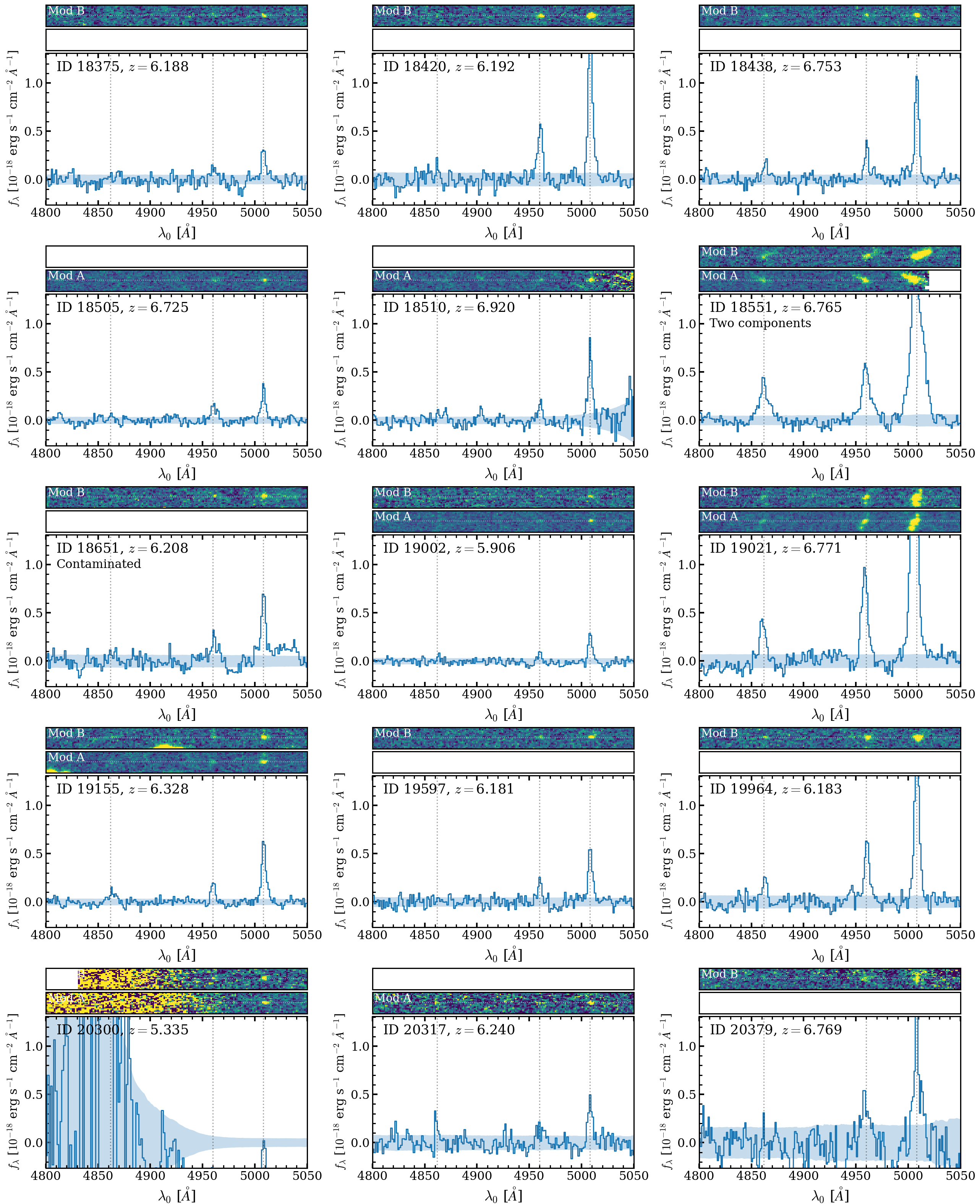}
    \caption{Fig. $\ref{fig:spec1}$, continued.}
    \label{fig:spec8}
\end{figure*}

In Figures $\ref{fig:spec1}$ - $\ref{fig:spec8}$ we show the 1D and 2D grism spectra of all 117 [OIII] emitters identified and studied in this paper. 

\end{CJK*}
\end{document}